\documentclass[english,showpacs,floatfix,preprint,10pt]{revtex4}
\usepackage{bbm}
\usepackage{mathrsfs}

\usepackage[dvips]{graphicx}
\usepackage{longtable}
\usepackage{amsmath,amssymb}
\usepackage{dcolumn}
\usepackage{latexsym}
\usepackage{babel}
\usepackage[latin1]{inputenc}
\usepackage{color}
\usepackage[colorlinks]{hyperref}
\begin{document}

\title{Effects of the interaction between dark energy and dark matter on cosmological parameters}
\author{Jian-Hua He, Bin Wang\footnote{E-mail address:
wangb@fudan.edu.cn}} \affiliation{Department of Physics, Fudan
University, Shanghai 200433, P. R. China}

\begin{abstract}
We examine the effects of possible phenomenological interactions
between dark energy and dark matter on cosmological parameters and
their efficiency in solving the coincidence problem. We work with
two simple parameterizations of the dynamical dark energy equation
of state and the constant dark energy equation of state. Using
observational data coming from the new 182 Gold type Ia supernova
samples, the shift parameter of the Cosmic Microwave Background
given by the three-year Wilkinson Microwave Anisotropy Probe
observations, and the baryon acoustic oscillation measurement from
the Sloan Digital Sky Survey, we perform a statistical joint
analysis of different forms of phenomenological interaction between
dark energy and dark matter.

\end{abstract}


\maketitle

\section{Introduction}
Our universe is believed undergoing an accelerated expansion
driven by a yet unknown dark energy (DE) \cite{1}-\cite{b310}.
Much effort has been devoted to understand the nature and the
origin of the DE. The leading interpretation of such a DE is a
cosmological constant with equation of state (EoS) $w=-1$.
Although this interpretation is consistent with observational
data, at the fundamental level it fails to be convincing. The
vacuum energy density falls far below the value predicted by any
sensible quantum field theory, and it unavoidably leads to the
coincidence problem, i.e., ``why are the vacuum and matter energy
densities of precisely the same order today?". To overcome the
coincidence problem, some sophisticated dynamical DE models either
relating the DE to a scalar field called quintessence with $w>-1$,
or to an exotic field called phantom with $w < -1$ \cite{2} have
been put forward to replace the cosmological constant.  But it is
doubtful that there is a clear winner in sight to explain the DE
and solve the coincidence problem.

DE contributes a significant fraction of the content of the
universe, it is natural to consider its interaction with the
remaining fields of the Standard Model in the framework of field
theory. The possibility that DE and DM can interact has been studied
in \cite{45}$^{-}$\cite{414}, among others. It has been shown that
the coupling between a DE (or quintessence) field and DM can provide
a mechanism to alleviate the coincidence problem \cite{45,410}. A
suitable choice of the coupling, motivated by holographic arguments,
can also lead to the crossing of the phantom barrier which separates
models with EoS $w> -1$ from models with $w <-1$ \cite{411,WLA06}.
In addition, it has been argued that an appropriate interaction
between DE and DM can influence the perturbation dynamics and affect
the lowest multipoles of the CMB spectrum \cite{412,415}. Arguments
using structure formation of galaxies suggested that the strength of
the coupling could be as large as the QED fine structure constant
\cite{412,416}. More recently, it was shown that such an interaction
could be inferred from the expansion history of the universe, as
manifested in, e.g., the supernova data together with CMB and
large-scale structure\cite{Feng07}. Nevertheless, the observational
limits on the strength of such an interaction remain weak
\cite{418}. In addition, it was suggested that the dynamical
equilibrium of collapsed structures would be affected by the
coupling of DE to DM \cite{426,427}. The basic idea is that the
virial theorem is distorted by the non-conservation of mass caused
by the coupling. This problem has been analyzed precisely in
\cite{Abdalla07}.

The interaction between DE and DM could be a major issue to be
confronted in studying the physics of DE. However, due to the
nature of these two components remain unknown, it will not be
possible to derive the precise form of the interaction from first
principles. One has to assume a specific coupling from the outset
\cite{L10, L11} or determine it from phenomenological requirements
\cite{L6}. In view of the continuity equations, the interaction
between DE and DM must be a function of the energy densities
multiplied by a quantity with units of inverse of time. For the
latter the obvious choice is the Hubble factor $H$. Thus, the
interaction between DE and DM could be expressed
phenomenologically in forms such as $Q = Q( H
\rho_{DM})$\cite{418, amendola}, $Q = Q(H \rho_{DE})$\cite{226},
$Q = Q(H (\rho_{DE}+\rho_{DM}))$\cite{49}-\cite{412} or most
generally in the form $Q = Q( H \rho_{DE}\, , H \rho_{DM})$ which
leads $Q \simeq \delta_1\, H \rho_{DE} + \delta_2\, H \rho_{DM}$
from the first term in the power law expansion. Considering the
couplings are terms in the lagragian which mix both DE and DM, one
may further presume that they could be parameterized by some
product of the densities of DE and DM, for example $Q=\lambda
\rho_{DE}\rho_{DM}$ \cite{MPLA}. Besides these phenomenological
descriptions of the interaction between DE and DM, recently there
is an attempt to describe the coupling from the thermodynamical
consideration\cite{LL,226}.

It is of great interest to investigate effects of the interaction
between DE and DM on the universe evolution, especially the
influence of different forms of the interaction on cosmological
parameters. This is the main motivation of the present paper.
Using the new 182 Gold type Ia supernova samples (SNIa), the
baryon acoustic oscillation measurement from the Sloan Digital Sky
Survey (SDSS-BAO) and the shift parameter of the Cosmic Microwave
Background given by the three-year Wilkinson Microwave Anisotropy
Probe observations (CMB shift), we are going to examine the
efficiency of different coupling forms in solving the cosmic
coincidence problem. In our study we will not specify any special
model of DE. Considering recent accurate data analysis showing
that the time varying DE gives a better fit than a cosmological
constant and in particular, DE EoS can cross $-1$ around $z = 0.2$
from above to below\cite{Y9}, we will employ two commonly used
parameterizations in our work, namely
$\omega_I=\omega_0+\omega_1z/(1+z)=\omega_0+\omega_1(1-a)$ and
$\omega_{II}=\omega_0+\omega_1z/(1+z)^2=\omega_0+(a-a^2)\omega_1$.
For comparison we will also examine the effects of the interaction
between DE and DM for the constant EoS of DE.

\section{Phenomenological models describing the interaction between DE and DM}
In the FRW universe, the Friedmann equation reads
\begin{equation}
3M_p^2H^2=\rho_r+\rho_b+\rho_{DM}+\rho_{DE}.
\end{equation}
Using the critical energy density $ \rho_{c}^0=3M_p^2H_0^2,$ we
can rewrite the Friedmann equation in the form
\begin{equation}
E^2(z)=\frac{H^2(z)}{H_0^2}=\Omega_b^0(1+z)^3+\Omega_r^0
(1+z)^4+\rho_{DE}(z)/\rho_{c}^0+\rho_{DM}(z)/\rho_{c}^0.
\end{equation}
The energy density of the radiation is given by,
\begin{equation}
\rho_r^0=\sigma_bT_{cmb}^4
\end{equation}
where $\sigma_b$ is the Stefan-Boltzmann constant,
$T_{cmb}=2.726K$ is the CMB temperature at present. $\rho_{DE}(z)$
represents the DE density and $\rho_{DM}$ is the energy density of
the cold DM.

When DE and DM interact with each other, we have
\begin{eqnarray}\label{consv}
\dot{\rho}_{DM}&+&3H\rho_{DM}= Q \, , \\
\label{convs2b} \dot{\rho}_{DE}&+&3H(1+w_{DE})\rho_{DE}= -Q \, ,
\end{eqnarray}
respectively, where $Q$ denotes the interaction term. Notice that
the overall energy density of dark sectors $\rho_{DM} +\rho_{DE}$
is conserved.

From the equations above, phenomenological forms of the
interaction between DE and DM must be a function of the energy
densities multiplied by a quantity with units of inverse of time.
The obvious choice of the quantity with units of inverse of time
is the Hubble factor $H$. The energy densities could be chosen as
density of either DE, DM, DE plus DM, the linear combination of DE
and DM or the product of DE and DM.

We will compare these different phenomenological expressions of the
interaction between DE and DM and investigate their effects on
cosmological evolutions and efficiencies in solving the coincidence
problem.
\subsection{The coupling is proportional to the energy density of DE}
If the interaction term is proportional to the energy density of
DE,we choose $Q=\delta H\rho_{DE}$ and rewrite the continuity
equations~\ref{consv} and ~\ref{convs2b}
\begin{eqnarray}
\dot{\rho}_{DM}+3H\rho_{DM}-\delta H \rho_{DE}=0\label{cpDEa}\\
\dot{\rho}_{DE}+3H\rho_{DE}(1+\omega)+\delta H
\rho_{DE}=0\label{cpDEb}
\end{eqnarray}
Employing the first commonly used parameterization of the EoS
$\omega_I$, we obtain a closed form solution for (\ref{cpDEb})
\begin{equation}
\rho_{DE}=\rho_{DE}^0e^{-3\omega_1}e^{3\omega_1/(1+z)}(1+z)^{3(1+\omega_0+\omega_1)+\delta}.
\end{equation}
Insert this solution into Eq(\ref{cpDEa}), we get
\begin{equation}
\rho_{DM}=(1+z)^3\rho_{DM}^0-\delta
\rho_{DE}^0(1+z)^3\int_0^ze^{3\omega_1/(1+z)-3\omega_1}(1+z)^{3(\omega_0+\omega_1)+\delta-1}dz.
\end{equation}
Choosing the second parameterization of the EoS $\omega_{II}$, we
have
\begin{equation}
\rho_{DE}=\rho_{DE}^0e^{\frac{3}{2}\omega_1+\frac{3}{2}\omega_1/(1+z)^2-3\omega_1/(1+z)}(1+z)^{3(1+\omega_0)+\delta},
\end{equation}
and
\begin{eqnarray}
\rho_{DM} & = &
\rho_{DM}^0(1+z)^3-\delta\rho_{DE}^0(1+z)^3e^{\frac{3}{2}\omega_1}\int_0^ze^{\frac{3}{2}\omega_1/(1+z)^2-3\omega_1/(1+z)}(1+z)^{\delta+3\omega_0-1}dz.
\end{eqnarray}
\subsection{The coupling is proportional to the energy density of DM}
 Expressing the interaction between DE and DM in proportional to
the DM energy density $Q=\delta H\rho_{DM}$,we can rewrite the
continuity equations as
\begin{eqnarray}
\dot{\rho}_{DM}+3H\rho_{DM}-\delta H \rho_{DM}=0,\label{cpDMa}\\
\dot{\rho}_{DE}+3H\rho_{DE}(1+\omega)+\delta H \rho_{DM}=0.
\label{cpDMb}
\end{eqnarray}
Employing $\omega_I$ as the parameterization of the EoS,  from
Eq(\ref{cpDMa}) we obtain,
\begin{equation}
\rho_{DM}=\rho_{DM}^0 (1+z)^{3-\delta}.
\end{equation}
Insert this $\rho_{DM}$ into Eq(\ref{cpDMb}), we arrive at
\begin{eqnarray}
\rho_{DE} & = &
(1+z)^{3(1+\omega_0+\omega_1)}e^{3\omega_1/(1+z)}\delta\rho_{DM}^0
\int_0^z
e^{-3\omega_1/(1+z)}(1+z)^{-3(\omega_0+\omega_1)-\delta-1}dz \\
& &
+\rho_{DE}^0(1+z)^{3(1+\omega_0+\omega_1)}e^{3\omega_1/(1+z)-3\omega_1}.
\end{eqnarray}
Using $\omega_{II}$ as the EoS, we find the evolution equation for
DM as
\begin{equation}
\rho_{DM}=\rho_{DM}^0 (1+z)^{3-\delta}.
\end{equation}
Then from Eq(\ref{cpDMb}) we have the evolution of the DE reads
\begin{eqnarray}
\rho_{DE}
 & = &(1+z)^{3(1+\omega_0)}e^{-\frac{3\omega_1}{1+z}+\frac{3}{2}\frac{\omega_1}{(1+z)^2}}\delta
\rho_{DM}^0\int_0^z(1+z)^{-3\omega_0-\delta-1}e^{\frac{3\omega_1}{1+z}-\frac{3}{2}\frac{\omega_1}{(1+z)^2}}dz\\
& &
+(1+z)^{3(1+\omega_0)}e^{-\frac{3\omega_1}{1+z}+\frac{3}{2}\frac{\omega_1}{(1+z)^2}}e^{\frac{3}{2}\omega_1}\rho_{DE}^0.
\end{eqnarray}
\subsection{The  coupling is proportional to the total energy density of DE and
DM} We can also write the interaction between DE and DM in
proportion to the total energy densities of DE and DM, so that the
continuity equations read
\begin{eqnarray}
\dot{\rho}_{DM}+3H\rho_{DM}-\delta H(\rho_{DM}+\rho_{DE})=0,\label{totala}\\
\dot{\rho}_{DE}+3H\rho_{DE}(1+\omega)+\delta H
(\rho_{DM}+\rho_{DE})=0.\label{totalb}
\end{eqnarray}
The above equations can be converted to
\begin{eqnarray}
\frac{d\rho_{DM}}{dz}-\frac{3}{1+z}\rho_{DM}+\frac{\delta}{1+z}(\rho_{DE}+\rho_{DM})=0,\label{a}\\
\frac{d\rho_{DE}}{dz}-\frac{3}{1+z}(1+\omega)\rho_{DE}-\frac{\delta}{1+z}(\rho_{DE}+\rho_{DM})=0.\label{b}
\end{eqnarray}
\subsection{The coupling is proportional to the linear combination of energy densities of DE and DM}
The phenomenological coupling between DE and DM could be expressed
most generally in the form  $Q = Q( H \rho_{DE}\, , H \rho_{DM})$
which leads $Q \simeq \delta_1\, H \rho_{DE} + \delta_2\, H
\rho_{DM}$ from the first term in the power law expansion. Thus the
continuity equations read
\begin{eqnarray}
\dot{\rho}_{DM}+3H\rho_{DM}- H(\delta_1\rho_{DE}+\delta_2\rho_{DM})=0,\label{totala}\\
\dot{\rho}_{DE}+3H\rho_{DE}(1+\omega)+H(\delta_1\rho_{DE}+\delta_2\rho_{DM})=0,\label{totalb}
\end{eqnarray}
which can be changed into
\begin{eqnarray}
\frac{d\rho_{DM}}{dz}-\frac{3}{1+z}\rho_{DM}+\frac{1}{1+z}(\delta_1\rho_{DE}+\delta_2\rho_{DM})=0,\label{a}\\
\frac{d\rho_{DE}}{dz}-\frac{3}{1+z}(1+\omega)\rho_{DE}-\frac{1}{1+z}(\delta_1\rho_{DE}+\delta_2\rho_{DM})=0.\label{b}
\end{eqnarray}
For the last two cases of describing the interaction between DE and
DM, it is not easy to write out analytic solutions of $\rho_{DE}$
and $\rho_{DM}$ when we substitute the parameterizations of EoS. We
will completely count on the numerical calculation to investigate
the time evolution of the DE and DM in these cases.
\subsection{The coupling is proportional to the product of energy densities of DE and DM}
The coupling can be considered in terms of the lagrangian which mix
both DE and DM, we can adopt the form of the interaction as the
product of the densities of DE and DM \cite{MPLA}, $Q=\lambda
\rho_{DE}\rho_{DM}$, where $\lambda=kM^{-3}$ and $M$ is considered
as a mass parameter and k is a dimensionless constant. Taking
$\Omega_x=\rho_x/\rho_{c0}$ and
$\rho_{c0}=3H_0^2M_p^2/(8\pi)=h^2\cdot 0.81\cdot 10^{-46} GeV^4$,
the dimensionless continuity equations for the DM and DE read
\begin{eqnarray}
  \frac{d\Omega_{DM}}{dz}-\frac{3}{1+z}\Omega_{DM}+\delta\frac{1}{(1+z)}\frac{\Omega_{DM}\Omega_{DE}}{\sqrt{\Omega_{DM}+\Omega_{DE}}}=0\nonumber \\
  \frac{d\Omega_{DE}}{dz}-\frac{3}{1+z}(1+\omega)\Omega_{DE}-\delta\frac{1}{(1+z)}\frac{\Omega_{DM}\Omega_{DE}}{\sqrt{\Omega_{DM}+\Omega_{DE}}}=0
\end{eqnarray}
where $\delta=\sqrt{3/(8\pi)}\lambda
\rho_{c0}^{1/2}M_p=\sqrt{3/(8\pi)}k(\frac{M_p}{M})(\frac{\rho_{c0}}{M^4})^{1/2}$,
which is a dimensionless coupling constant now.

So far we have listed all possible phenomenological descriptions of
the interaction between DE and DM. In the next section, we will
examine the viability of these descriptions by comparing with
observations.

\section{Observational constraints}
We will use the three-year WMAP (WMAP3) data\cite{GSpergel}, the
SN Ia data \cite{GRiess} and the Baryon Acoustic Oscillation (BAO)
measurement from the Sloan Digital Sky Survey \cite{GEisenstein}
to study the property of the interaction between DE and DM. In
\cite{GElgar}, the authors showed that combining the shift
parameters $R$ and the angular scale $l_a$ of the sound horizon at
recombination appears to be a good approximation of the full WMAP3
data. The model independent constraints on $R$ and $l_a$ was later
given \cite{GwangY} by using the WMAP3 data, where the covariance
matrix of the parameters $R, l_a$ and $\Omega_bh^2$ was also
provided. In the following we will use the shift parameter $R$,
the angular scale $l_a$ of the sound horizon at recombination and
their covariance matrix given in \cite{GwangY} and employ the
Monte-Carlo Markov Chain (MCMC) method to explore the parameter
space. Our MCMC code is based on the publicly available package
CMBeasy\cite{cmbeasy}.

For the SNIa data, we will calculate
$\chi^2=\Sigma_i\frac{[\mu_{obs}(zi) -\mu_(zi)]^2}{\sigma^2_i}$,
where the extinction-corrected distance modulus $\mu(z) = 5
log_{10}[d_L(z)/Mpc] + 25$, $\sigma_i$ is the total uncertainty in
the SN Ia data, and the luminosity distance is
$d_L(z)=(1+z)/H_0\int_0^{z_i}dz'/E(z')$ and the nuisance parameter
$H_0$ is marginalized over with flat prior.

For the SDSS data, we add the BAO parameter \cite{GEisenstein,
GSpergel},
\begin{equation}
\label{para1}
A=\frac{\sqrt{\Omega_{m}^0}}{E(0.35)^{1/3}}\left[\frac{1}{0.35}\int_0^{0.35}
\frac{dz'}{E(z')}\right]^{2/3}=0.469(0.95/0.98)^{-0.35}\pm
0.017,\nonumber
\end{equation}
to $\chi^2$.

For WMAP3 data, we first add the shift parameter\cite{GwangY}
\begin{equation}
\label{shift1}
R=\sqrt{\Omega_{m}^0}\int_0^{z_{ls}}\frac{dz'}{E(z')}=1.70\pm
0.03,\nonumber
\end{equation}
to $\chi^2$ at $z=1089\pm 1$. It was argued in \cite{GElgar,
GwangY} that the combination of the shift parameter and the
angular scale of the sound horizon at recombination gives much
better constraints on cosmological parameters. Thus we add the
angular scale of the sound horizon at recombination
\begin{equation}
\label{csla} l_a=\frac{\pi
R/\sqrt{\Omega_m^0}}{\int_{z_{ls}}^\infty dz c_s/E(z)}=302.2\pm
1.2,\nonumber
\end{equation}
where the sound speed $c_s=1/\sqrt{3(1+R_ba)}, R_b = 315000
\Omega_bh^2(T_{cmb}/2.7K)^{-4}$, $a$ is the scale factor, and
$\Omega_b h^2 = 0.022_{-0.00082}^{+0.00082}$.

We use the MCMC method to explore the parameter space. We treat
$H_0$ as also a fitting parameter, and we impose a prior of $H_0 =
72 \pm 8 km/s/Mpc$\cite{GFreedman}. By using two parameterizations
for the EoS of DE $\omega_I, \omega_{II}$, we obtain the parameter
space for different couplings respectively and we will show our
results in the following sections.For comparisons we will also
present the result for  constant DE EoS.
\subsection{The coupling is proportional to the energy density of DE}
\begin{figure}
\begin{center}
  \begin{tabular}{cc}
\includegraphics[width=2.0in,height=2.0in,angle=-90]{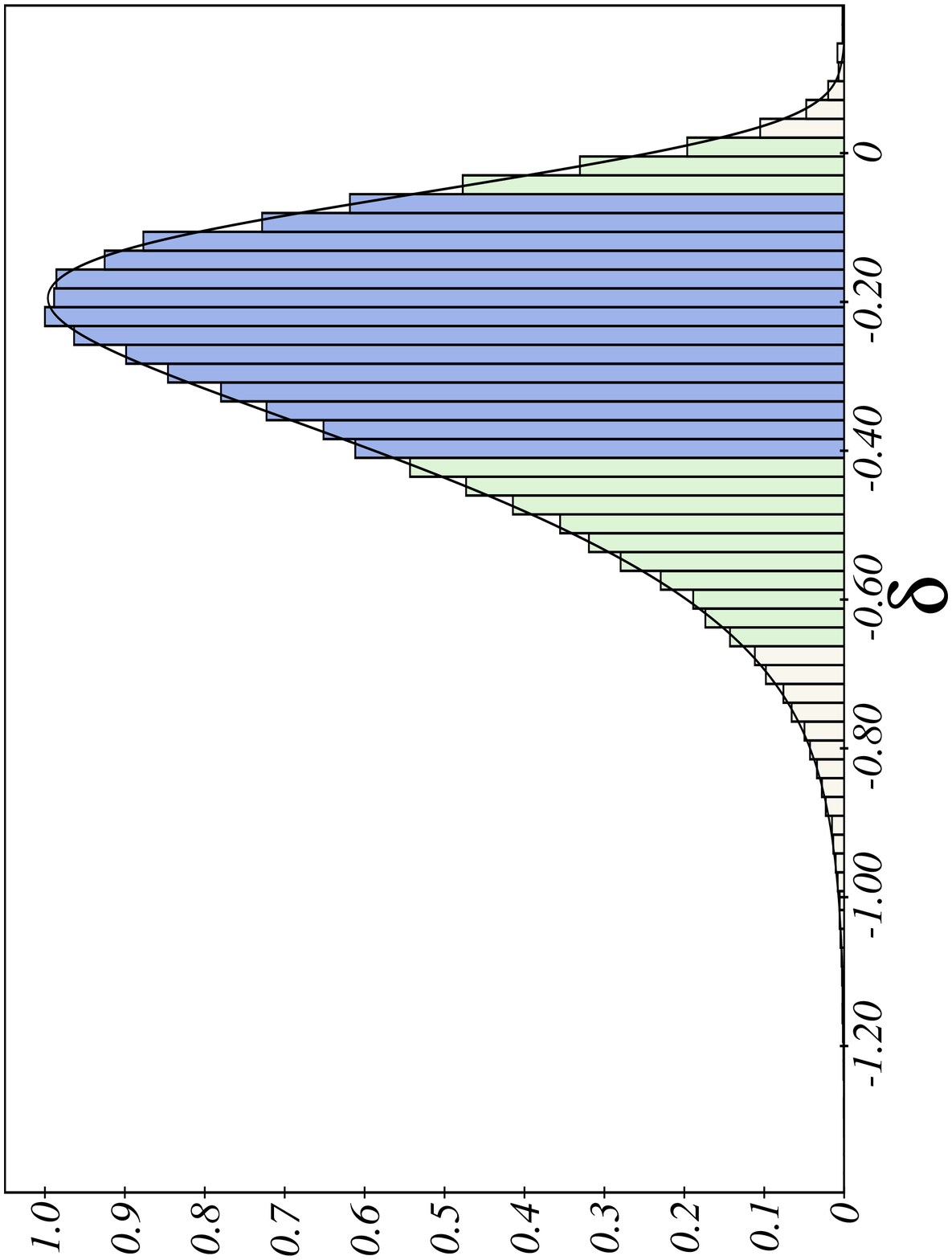}&
\includegraphics[width=2.0in,height=2.0in,angle=-90]{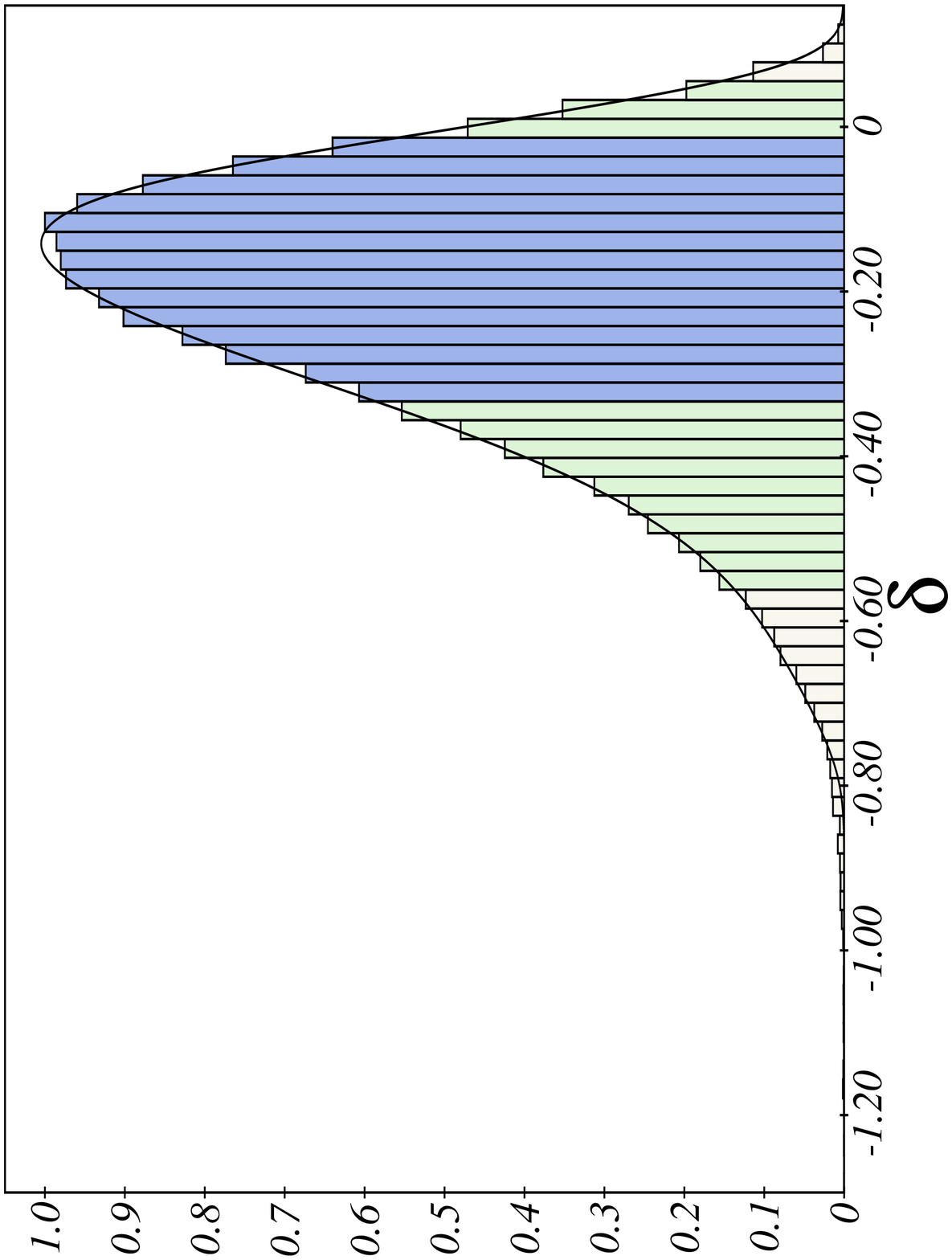}\nonumber \\
    (a)$\omega_I$&(b)$\omega_{II}$
  \end{tabular}
\end{center}
\caption{The likelihood for coupling parameters when the interaction
is chosen as $\delta H\rho_{DE}$}\label{figlikeDE}
\end{figure}

\begin{figure}
\begin{center}
  \begin{tabular}{ccc}
   \includegraphics[width=2in,height=2in]{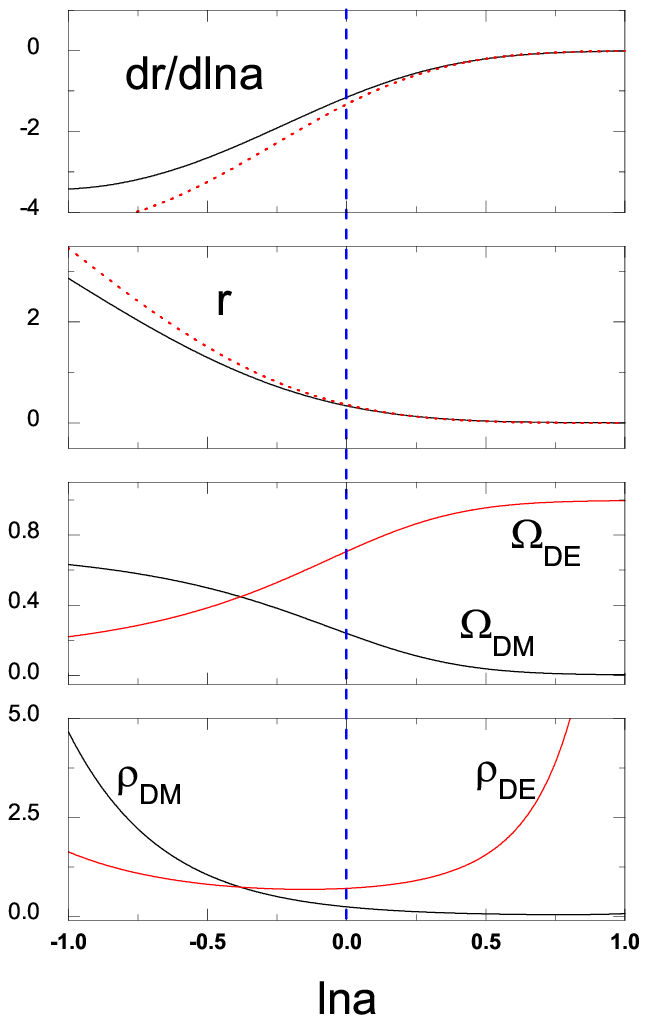}&
   \includegraphics[width=2in,height=2in]{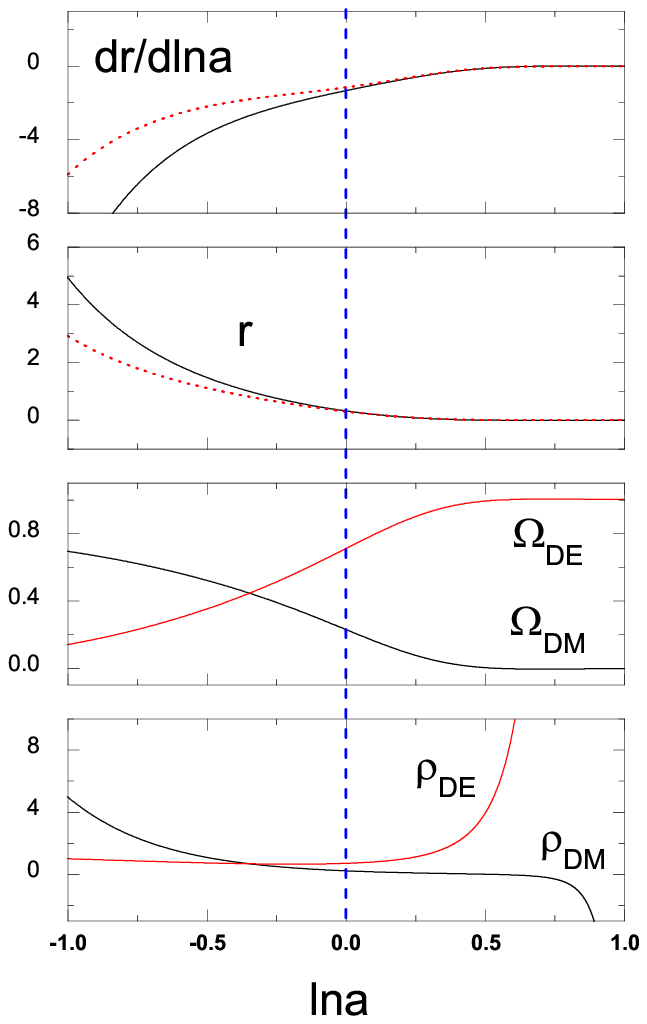}&\includegraphics[width=2in,height=2in]{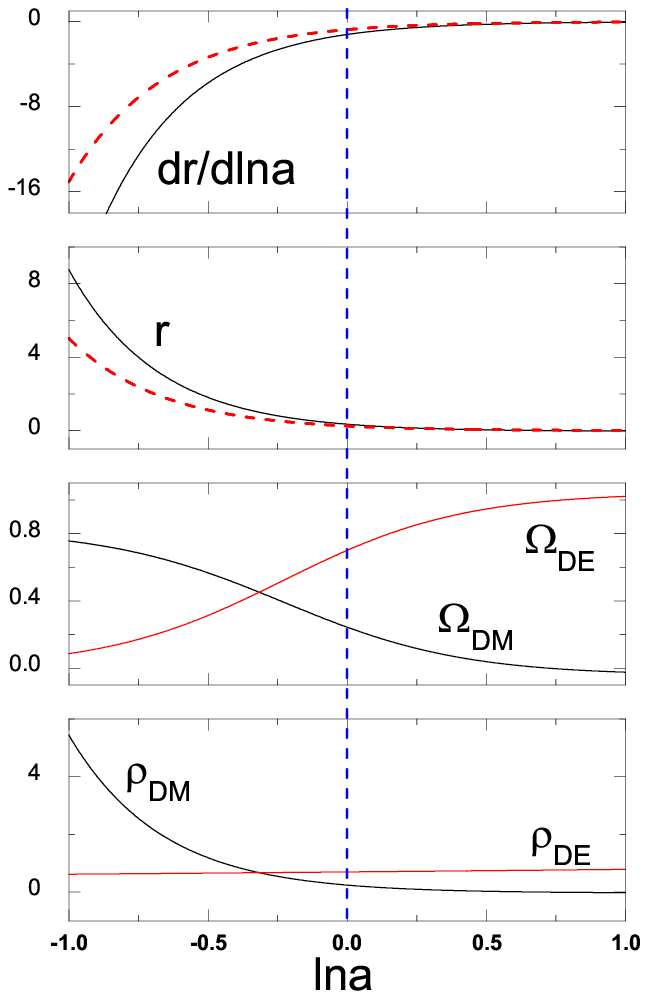}\\
   (a)$\omega_I$&(b)$\omega_{II}$&(c)constant $\omega$
  \end{tabular}
\end{center}
\caption{These figures show clearly the behaviors of different dark
sectors during the late time evolution of universe with best-fitted
data for coupling $\delta H\rho_{DE}$. The red dotted line denotes
the noninteracting case.The left one corresponds to EoS
$\omega_I$,the middle one is EoS $\omega_{II}$ and the right one is
constant EoS }\label{figDE}
\end{figure}

Choosing the first parameterization of the DE EoS $\omega_I$ and
expressing the coupling between DE and DM in proportional to the
energy density of DE, the results are shown in
Fig~\ref{figlikeDE}a and Fig~\ref{figDE}a. The quantity of
interest for analyzing the coincidence problem is the ratio
$r=\rho_{DM}/\rho_{DE}$ and its evolution with time. The behavior
of $dr/d\ln a$ deserves attention with regard to the coincidence
problem. Clearly we see that the positive coupling obtained from
the best fit data leads slower change of $r$ which makes the
coincidence problem less acute when compared with the case without
interaction shown by the red dotted line in Fig~\ref{figDE}a.When
we choose DE EoS to be $\omega_{II}$ and constant,results are
shown in Fig~\ref{figDE}b,c.From the behaviors of the ratio
$r=\rho_{DM}/\rho_{DE}$,for the constant DE EoS or
$\omega_{II}$,this phenomenological interaction cannot help to
alleviate the coincidence problem. Besides, the negative coupling
shown in Fig~\ref{figlikeDE}b for DE EoS $\omega_{II}$ leads
negative $\rho_{DE}$in the early time.When DE EoS is constant
$\rho_{DE}$ is nearly zero even in the inflationary era,which
makes it hard to count on this $\rho_{DE}$to explain the
infation.What's worse,the $\rho_{DM}$ will be negative in the
future. All these unphysical consequences are brought by the
negative coupling from the fitting.
\begin{center}
\begin{tabular}{c|c|ccccccc}
 \hline
 \hline
  Coupling & EoS &  & $\Omega_b^0$ & $\omega_0$ & $\omega_1$ & $\Omega_{DM}^0$ & $h_0$ & $\delta$ \\
\hline
  $\delta H\rho_{DE}$ & $\omega_I$   & Mean &$0.052_{ -0.006}^{+0.006}$ & $-0.82_{-0.23}^{+0.30}$ & $-1.97_{-1.64}^{+0.99}$ & $0.27_{-0.03}^{+0.03}$ & $0.66_{-0.04}^{+0.05}$ & $-0.28_{-0.20}^{+0.13}$  \\
          &    &Best-fitted& $0.054$ & $-1.18  $ & $1.21$ & $0.24$ & $0.63$ & $0.03$ \\
 \hline
   & $\omega_{II}$ & Mean &$0.055_{-0.006 }^{+0.007}$ & $-1.00_{-0.37}^{+0.45}$ & $-0.72_{-3.68}^{+2.65}$ & $0.25_{-0.03}^{+0.03}$ & $0.64_{-0.05}^{+0.05}$ & $-0.22_{-0.18 }^{+0.12}$    \\
   &               &Best-fitted& $0.059$ & $-1.26$ & $2.03$ & $0.23$ & $0.62$ & $ -0.08$ \\
 \hline
&     &     &$\Omega_b^0$ &  $\omega$ & $\Omega_{DM}^0$ & $h_0$ &$\delta$&\\
 \hline
        & constant $\omega$ & Mean &$0.055_{ -0.006}^{+0.007}$ & $-1.05_{-0.13}^{+0.12}$ & $0.25_{-0.02}^{+0.02}$ & $0.64_{-0.04}^{+0.05}$ & $-0.19_{-0.13}^{+0.10}$ & \\
       &            & Best-fitted & $0.057$ & $-1.00  $ & $0.24$ & $0.63$ & $-0.13$ & \\
 \hline
 \hline
\end{tabular}
\end{center}

\subsection{The coupling is proportional to the energy density of DM}
\begin{figure}
\begin{center}
  \begin{tabular}{cc}
\includegraphics[width=2.0in,height=2.0in,angle=-90]{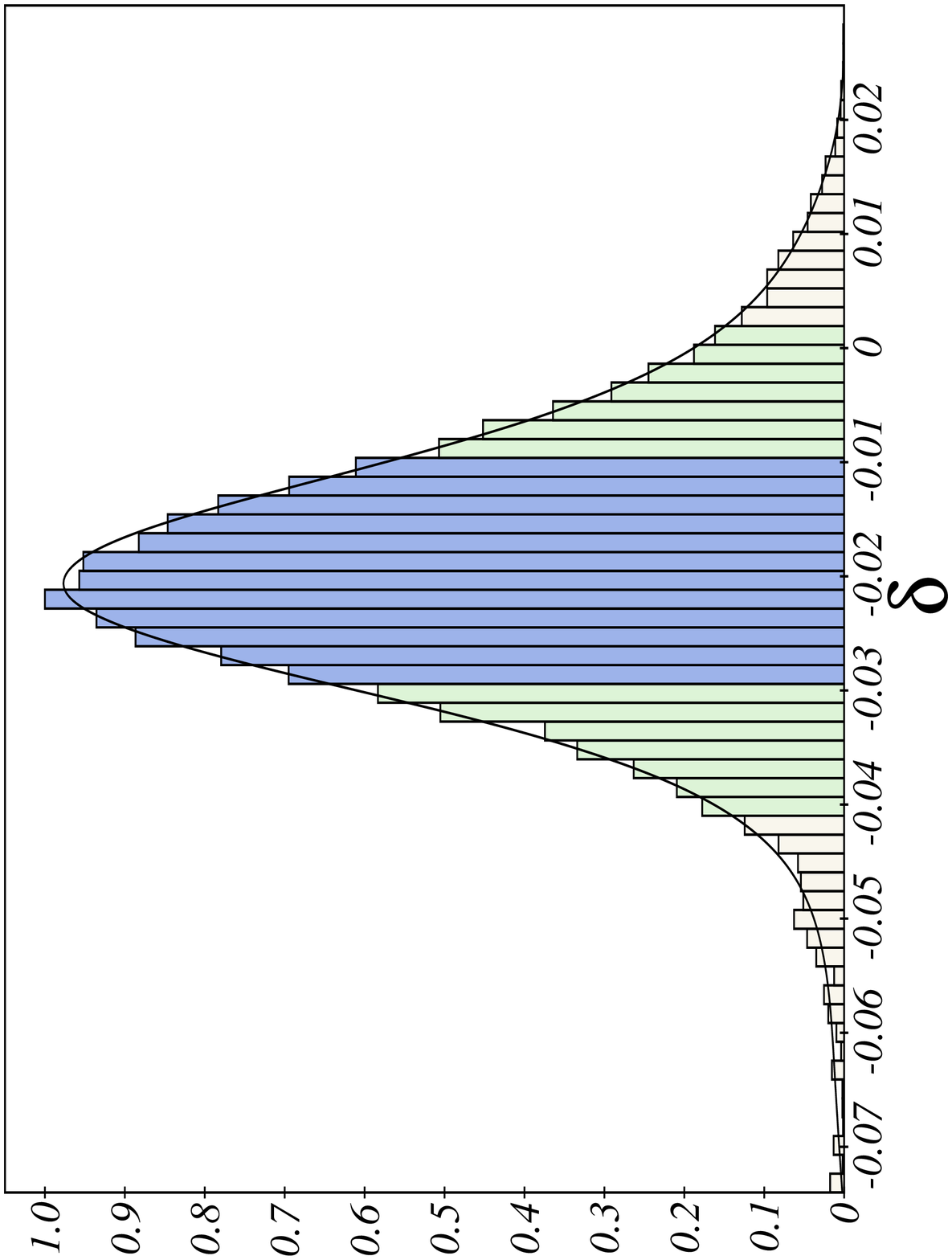}&
\includegraphics[width=2.0in,height=2.0in,angle=-90]{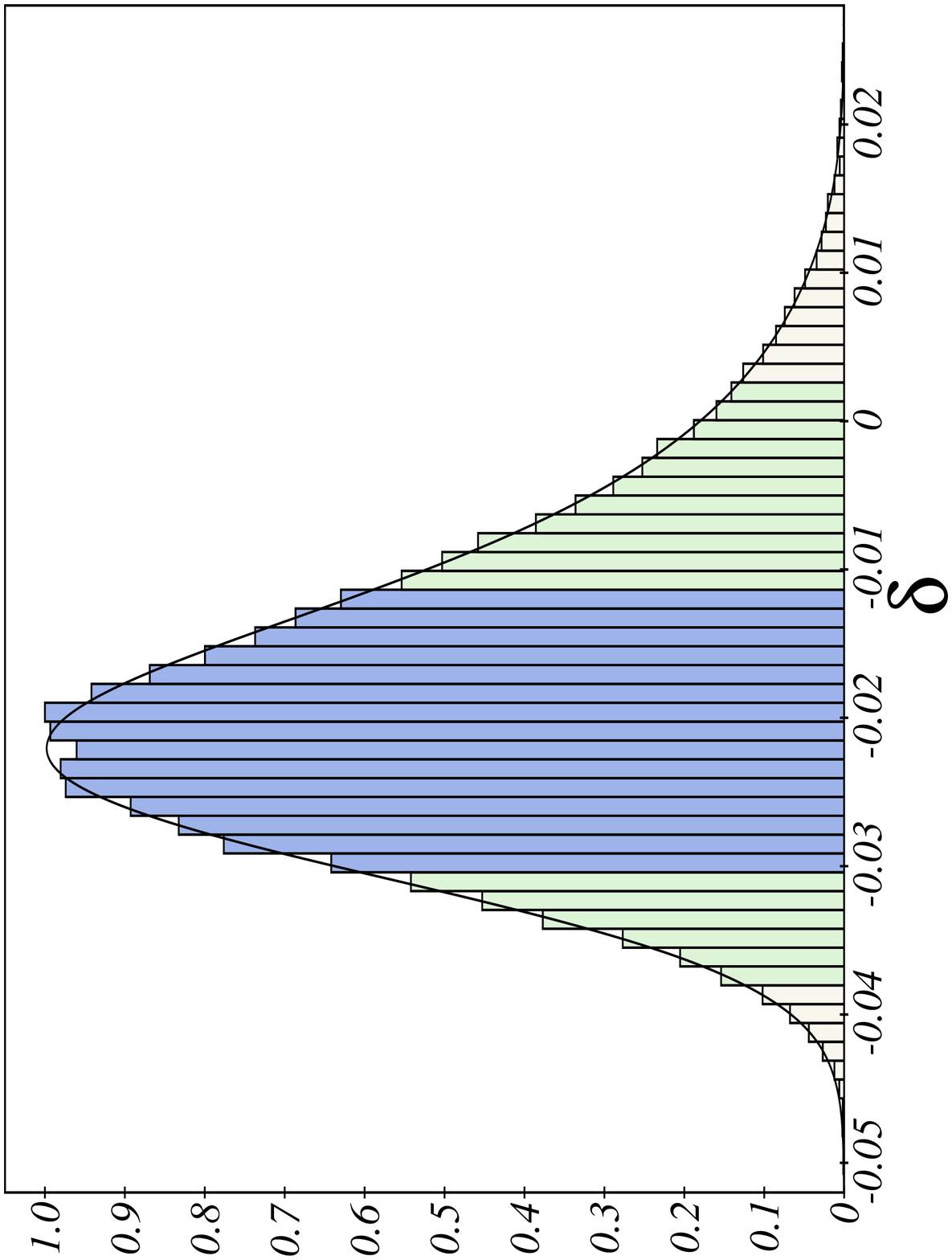}\nonumber \\
 (a)$\omega_I$&(b)$\omega_{II}$
  \end{tabular}
\end{center}
\caption{The likelihood for coupling parameters when the interaction
is chosen as $\delta H\rho_{DM}$.} \label{figlikeDM}
\end{figure}

\begin{figure}
\begin{center}
  \begin{tabular}{ccc}
   \includegraphics[width=2in,height=2in]{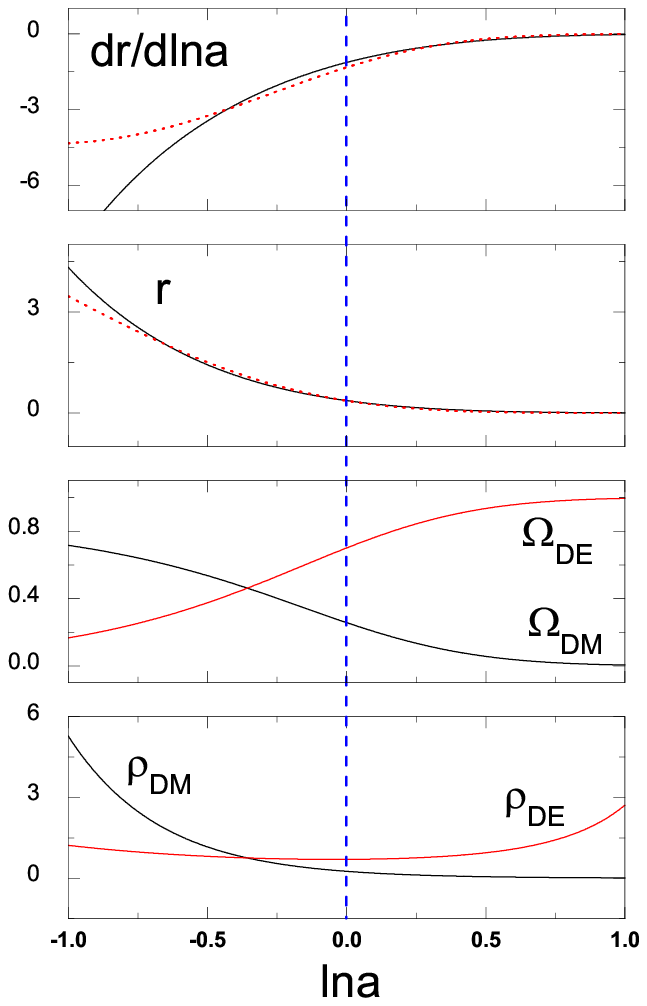}&
   \includegraphics[width=2in,height=2in]{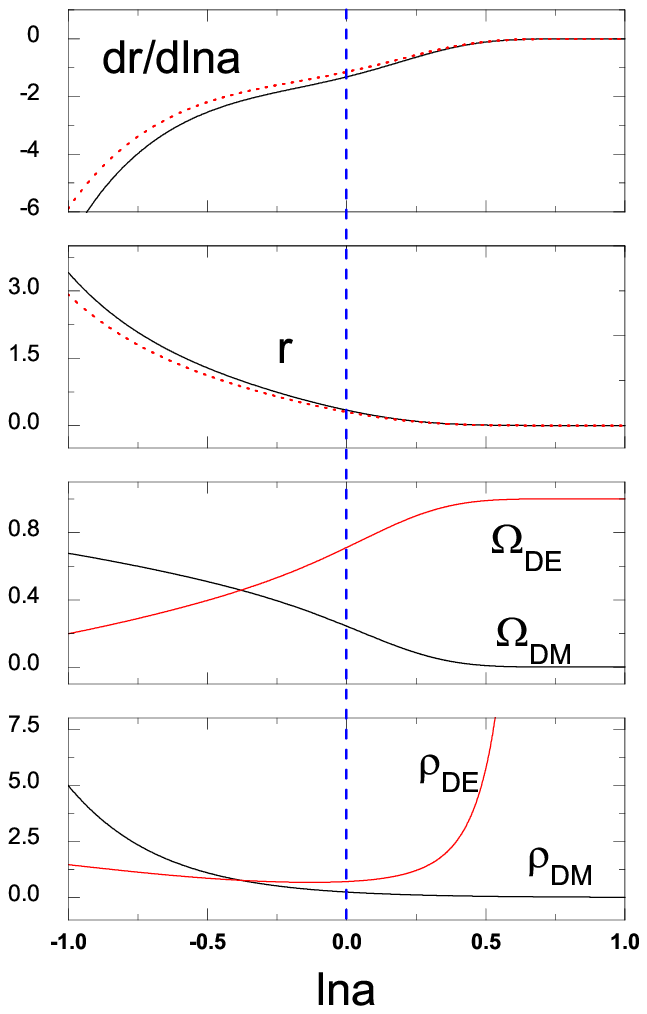}&\includegraphics[width=2in,height=2in]{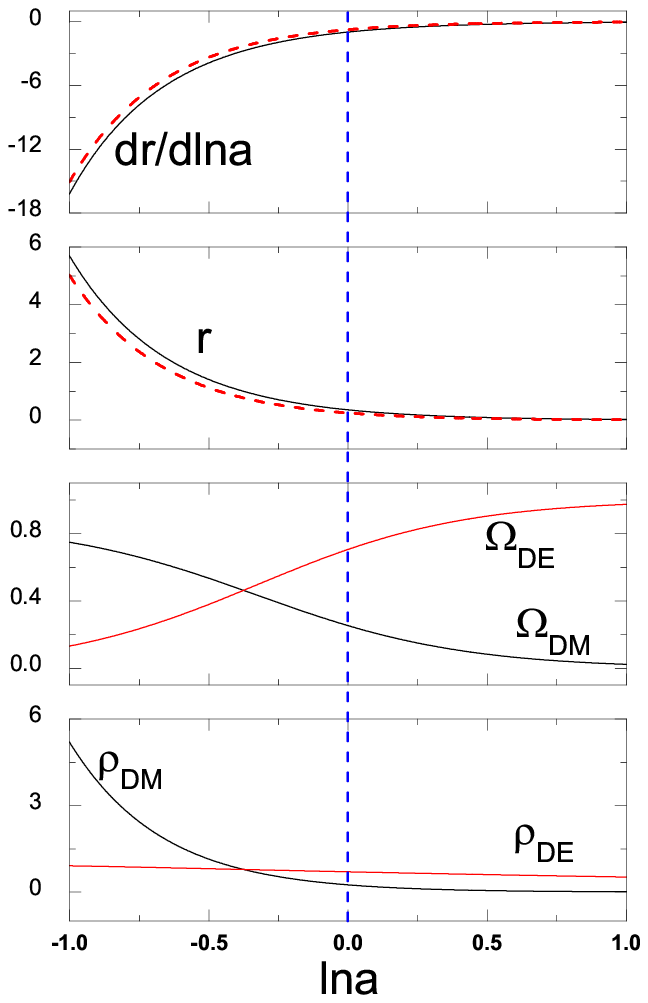}\\
   (a)$\omega_I$&(b)$\omega_{II}$&constant $\omega$
  \end{tabular}
\end{center}
\caption{These figures show clearly the behaviors of different
dark sectors during the late time evolution of universe with
best-fitted data for the coupling $\delta H\rho_{DM}$. The red
dotted line denotes the noninteracting data.The left one
corresponds to EoS $\omega_I$,the middle one is EoS $\omega_{II}$
and the right one is constant EoS }\label{figDM}
\end{figure}

Using the first parameterization of the DE EoS and choosing the
coupling between DE and DM in proportional to the energy density
of DM, we show the likelihood in Fig~\ref{figlikeDM}a and the
behaviors of cosmological parameters in Fig~\ref{figDM}a.
Comparing lines of $dr/d\ln a$ for the interacting and
noninteracting cases, it seems that near the present time the
interaction between DE and DM makes the change of $r$ a bit slower
than the noninteracting case, which makes the coincidence problem
less serious. However since the coupling is negative from the best
fit, we encounter the negative $\rho_{DE}$ in the early epoch of
the universe.This sounds unphysical\cite{amendola}, although in
some modified gravity theory negative value of $\rho_{DE}$ was
allowed\cite{amendola28}.The results for choosing DE EoS to be
$\omega_{II}$ and constant are shown in Fig~\ref{figDM}b and
Fig~\ref{figDM}c.It is clear from the behavior of $r$ that the
coincidence problem cannot be alleviated by choosing this
phenomenological interaction form when DE EoS are of $\omega_{II}$
or constant. Moreover, negative $\rho_{DE}$ at very early epoch
appears again.
\begin{center}
\begin{tabular}{c|c|ccccccc}
 \hline
 \hline
  Coupling & EoS &  & $\Omega_b^0$ & $\omega_0$ & $\omega_1$ & $\Omega_{DM}^0$ & $h_0$ & $\delta$ \\
   $\delta H\rho_{DM}$ & $\omega_I$ & Mean  & $0.047_{-0.008}^{+0.013}$ & $-0.99_{- 0.17}^{+ 0.20}$ & $0.34_{- 0.92}^{+ 0.37}$ & $0.25_{- 0.02}^{+ 0.02}$ & $0.69_{- 0.09}^{+ 0.08}$ & $-0.020 _{- 0.010}^{+ 0.010}$ \\
    &  & Best-fitted  & $0.044$ & $-1.02$ & $0.60$ & $0.26$ & $0.71$ & $-0.024$  \\
  \hline
   & $\omega_{II}$& Mean & $0.048_{- 0.008}^{+ 0.012}$ & $-1.25_{- 0.31}^{+ 0.31}$ & $2.35_{- 2.05}^{+ 1.96}$ & $0.25_{- 0.02}^{+ 0.02}$ & $0.69_{- 0.09}^{+ 0.08}$ & $-0.019_{- 0.008}^{+ 0.010}$ \\
   & & Best-fitted & $0.046$ & $-1.28 $ & $2.65$ & $ 0.24  $ & $0.69 $ & $-0.021$    \\
  \hline
&     &     &$\Omega_b^0$ &  $\omega$ & $\Omega_{DM}^0$ & $h_0$ &$\delta$&\\
\hline
      & constant $\omega$ & Mean  & $0.046_{-0.008}^{+0.010}$ & $-0.90_{- 0.08}^{+ 0.08}$ & $0.25_{- 0.02}^{+ 0.02}$ & $0.71_{- 0.08}^{+ 0.07}$ & $-0.018_{- 0.008}^{+ 0.010}$ & \\
      &              & Best-fitted  & $0.041$ & $-0.90$ & $0.25$ & $0.73$ & $-0.022$ &  \\
  \hline
  \hline
\end{tabular}
\end{center}
\subsection{The  coupling is proportional to the total energy density of DE and
DM} 
\begin{figure}
\begin{center}
\begin{tabular}{cc}
\includegraphics[width=2.0in,height=2.0in,angle=-90]{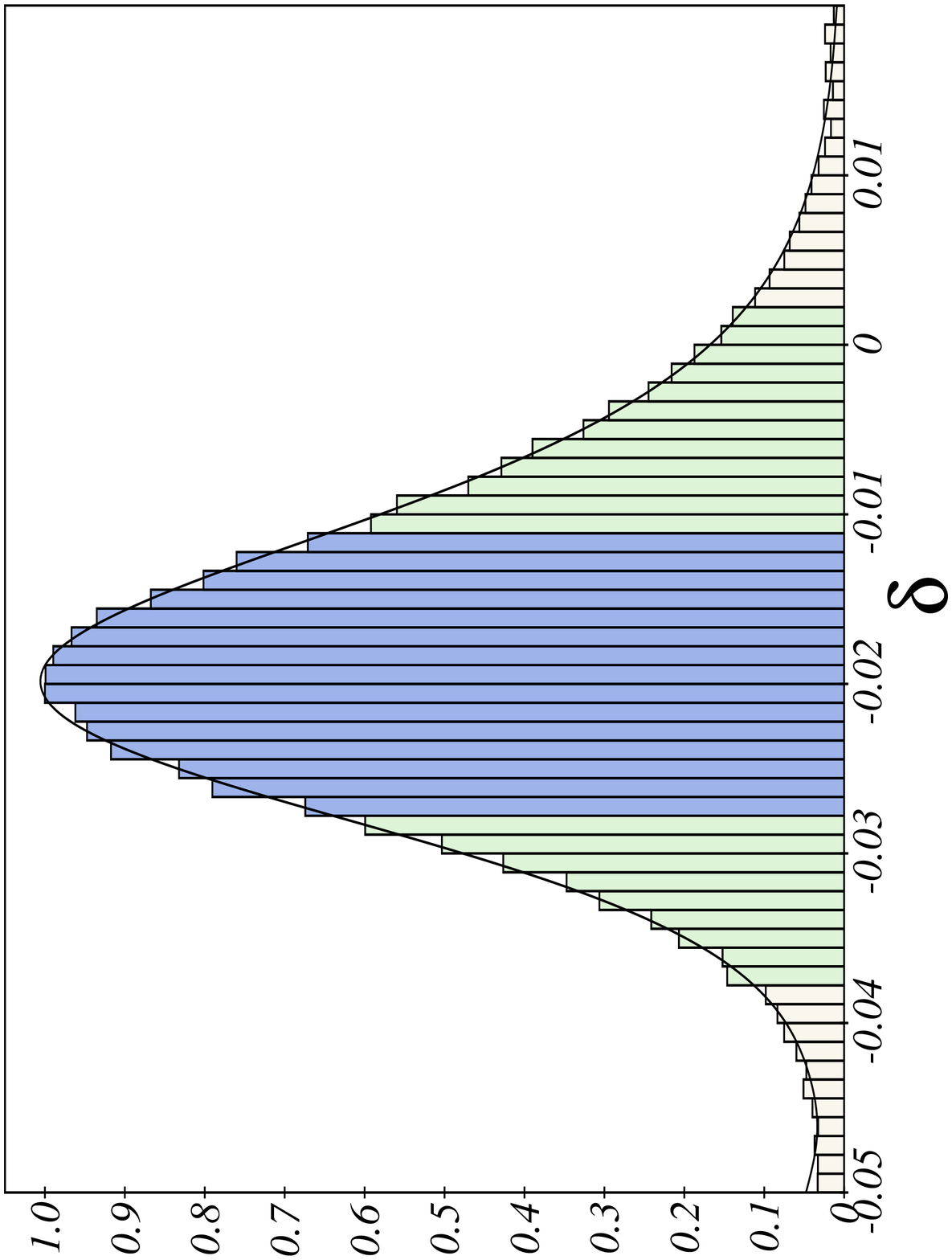}&
\includegraphics[width=2.0in,height=2.0in,angle=-90]{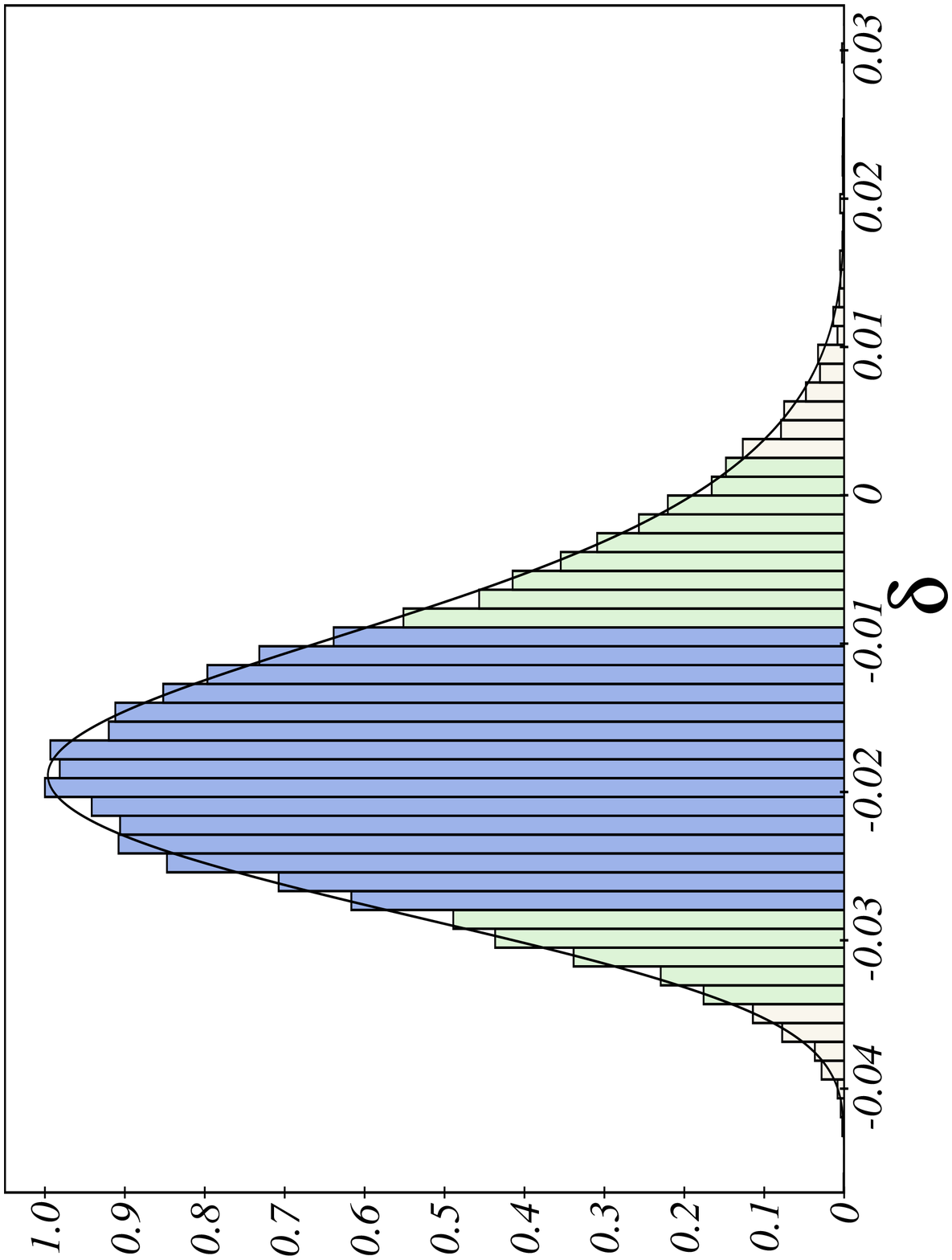}\nonumber\\
(a)$\omega_I$&(b)$\omega_{II}$
\end{tabular}
\end{center}
 \caption{The likelihood for
coupling parameters when the interaction is chosen as $\delta
H(\rho_{DM}+\rho_{DE})$.} \label{figlikeT}
\end{figure}

\begin{figure}
\begin{center}
  \begin{tabular}{ccc}
   \includegraphics[width=2in,height=2in]{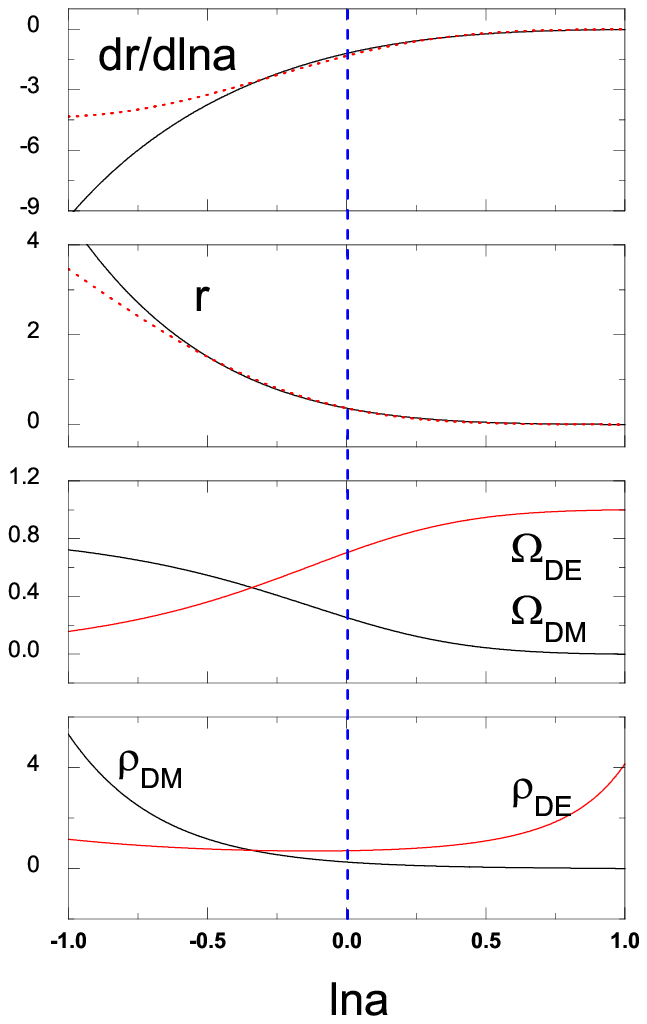}&
   \includegraphics[width=2in,height=2in]{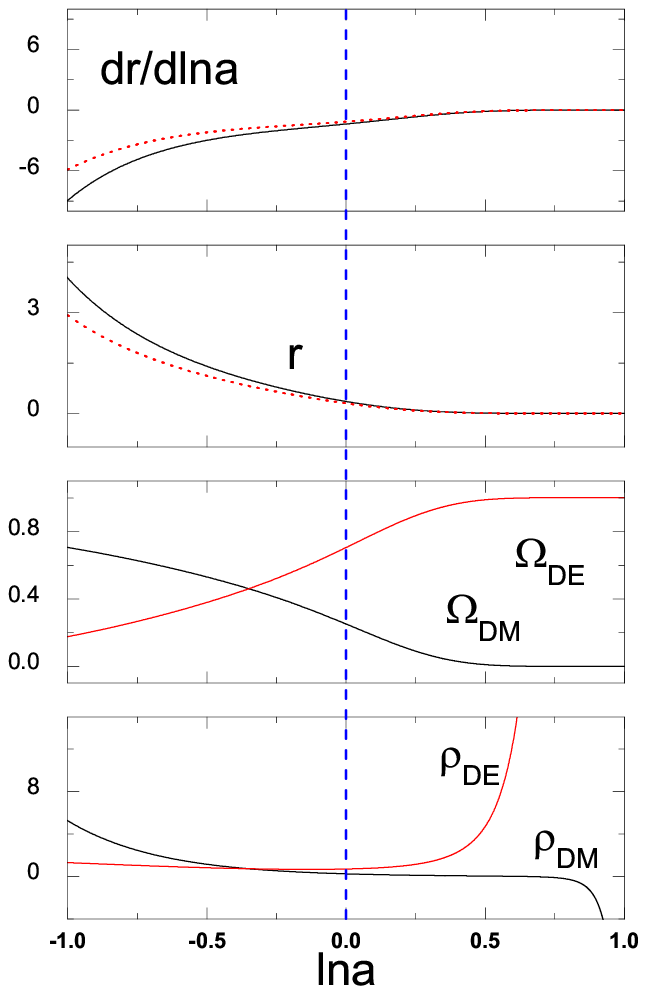}&\includegraphics[width=2in,height=2in]{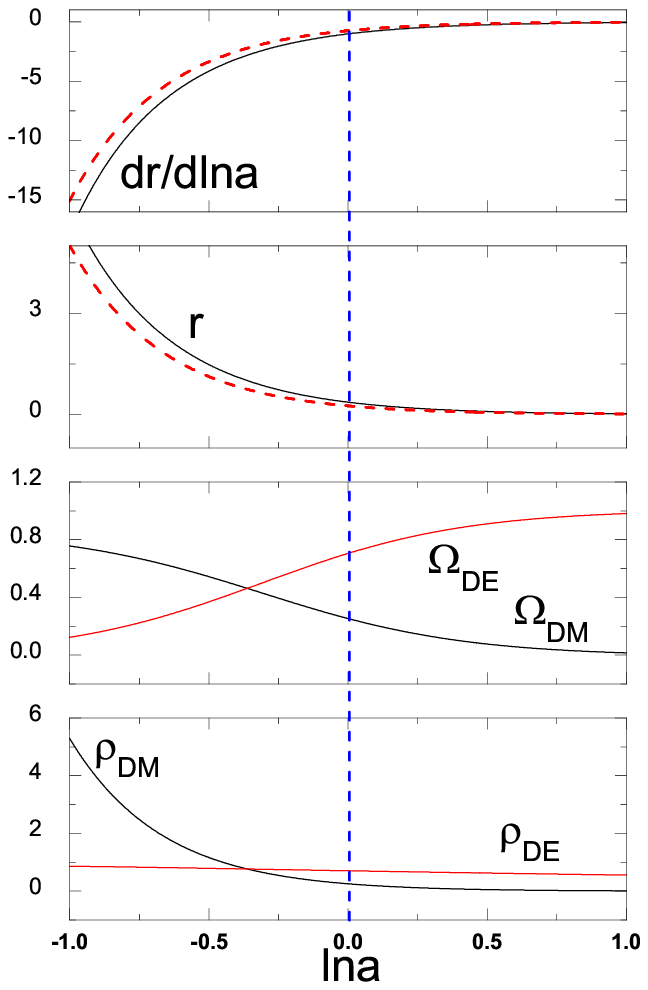}\\
   (a)$\omega_I$&(b)$\omega_{II}$&constant $\omega$
  \end{tabular}
\end{center}
\caption{These figures show clearly the behaviors of different
dark sectors during the late time evolution of universe with
best-fitted data for the coupling $\delta H(\rho_{DM}+\rho_{DE})$.
The red dotted line denotes the noninteracting case.The left one
corresponds to EoS $\omega_I$,the middle one is EoS $\omega_{II}$
and the right one is constant EoS }\label{figT}
\end{figure}

The results of the evolution of cosmological parameters for
employing $\omega_{I}$and expressing the interaction between DE
and DM in proportional to the total energy density of DE and DM
are shown in Fig~\ref{figlikeT}a and Fig~\ref{figT}a.In
Fig~\ref{figT}a we see that,$r$changes a bit slower than that of
the noninteracting case at very recent epoch,and this might help
to alleviate the coincidence problem. However in the early epoch
the negative coupling obtained from fitting makes $\rho_{DE}$
negative again.Choosing DE EoS to be $\omega_{II}$,besides the
negative $\rho_{DE}$ in the early time, $\rho_{DM}$ will appear
negative in the future. For constant DE EoS,$\rho_{DE}$ can aslo
be negative in the early epoch. Furthermore from
Fig~\ref{figlikeT}b,c, we see that the coincidence problem has not
been alleviated.
\begin{center}
\begin{tabular}{c|c|ccccccc}
 \hline
 \hline
  Coupling & EoS &  & $\Omega_b^0$ & $\omega_0$ & $\omega_1$ & $\Omega_{DM}^0$ & $h_0$ & $\delta$ \\
  \hline
  $\delta H(\rho_{DM}+\rho_{DE})$ &$\omega_I$ & Mean &$0.046_{- 0.008}^{+ 0.012}$ & $-0.98_{- 0.17}^{+ 0.20}$ & $0.21_{- 0.87}^{+ 0.49}$ & $0.25_{- 0.02}^{+ 0.02}$ & $0.70_{- 0.09}^{+ 0.08}$ & $-0.018 _{- 0.008}^{+0.009}$  \\
  && Best-fitted & $0.044$& $-1.07$& $0.71$ & $0.25$ & $0.71$ & $-0.022$  \\
  \hline
  &$\omega_{II}$& Mean &$0.048_{- 0.008}^{+ 0.013}$& $-1.22_{- 0.34}^{+ 0.34}$& $2.09_{- 2.24}^{+ 2.08}$ & $0.246_{- 0.02}^{+ 0.02}$ & $0.69_{- 0.09}^{+ 0.08}$ & $-0.017_{- 0.009}^{+ 0.010}$  \\
   &   & Best-fitted &  $0.044$& $-1.25 $& $2.39$ & $0.25$ & $0.71 $ & $-0.021  $  \\
\hline
&     &     &$\Omega_b^0$ &  $\omega$ & $\Omega_{DM}^0$ & $h_0$ &$\delta$&\\
\hline
    & constant $\omega$ & Mean &$0.045_{- 0.008}^{+ 0.010}$ & $-0.91_{- 0.08}^{+ 0.08}$ & $0.25_{- 0.02}^{+ 0.02}$ & $0.71_{- 0.08}^{+ 0.08}$ & $-0.017_{- 0.008}^{+ 0.009}$&  \\
    &             & Best-fitted & $0.042$& $-0.91$& $0.25$ & $0.72$ & $-0.020$&   \\
  \hline
  \hline
\end{tabular}
\end{center}

\subsection{The coupling is proportional to the linear combination of energy densities of DE and DM}
\begin{figure}
\begin{center}
  \begin{tabular}{ccc}
   \includegraphics[width=2in,height=2in]{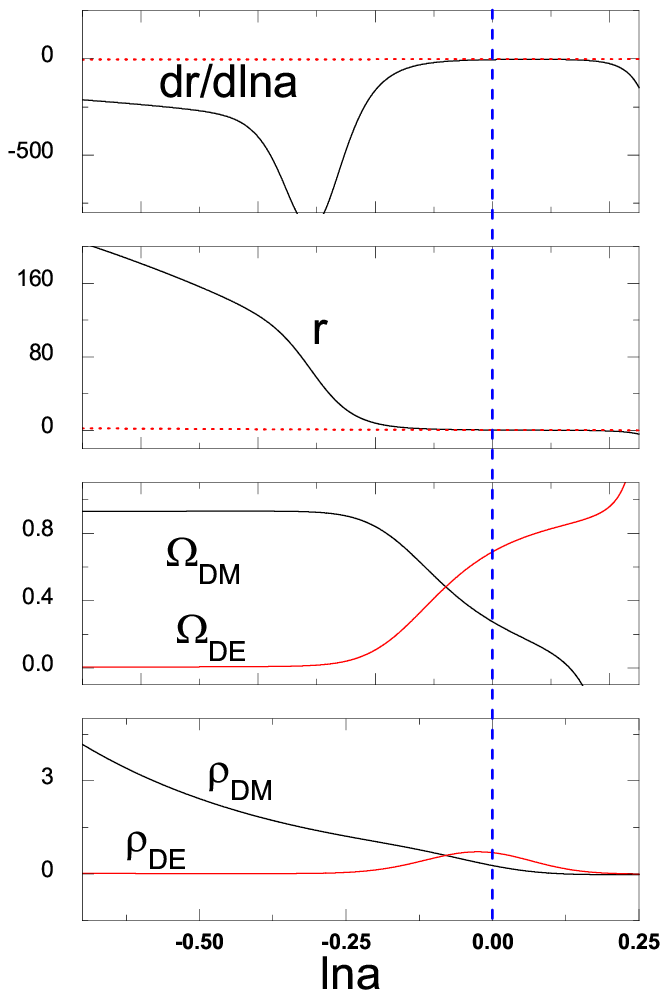}&
   \includegraphics[width=2in,height=2in]{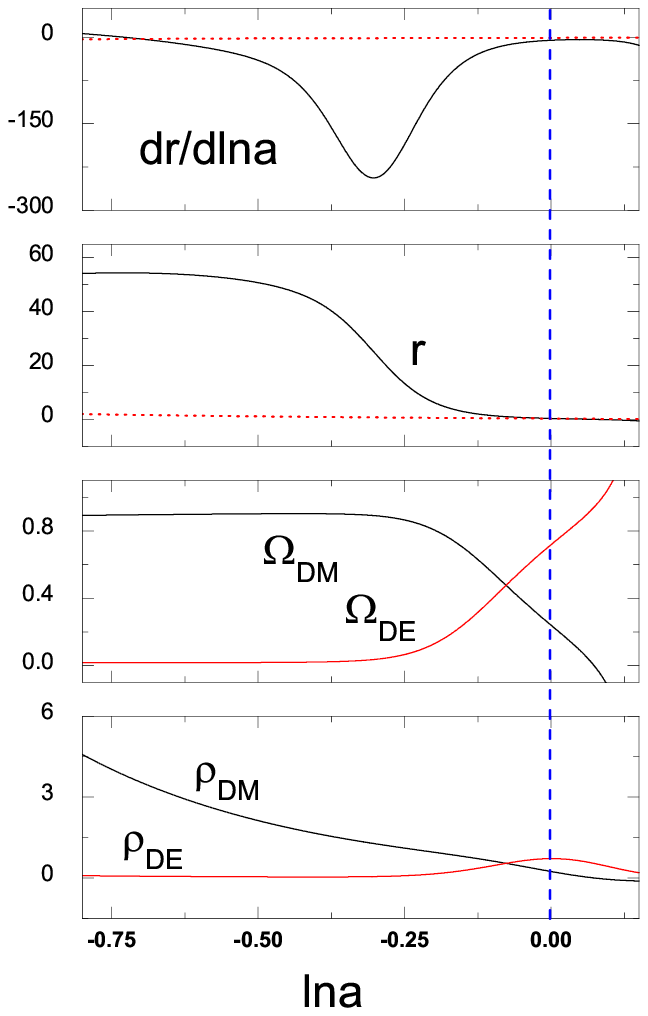}&\includegraphics[width=2in,height=2in]{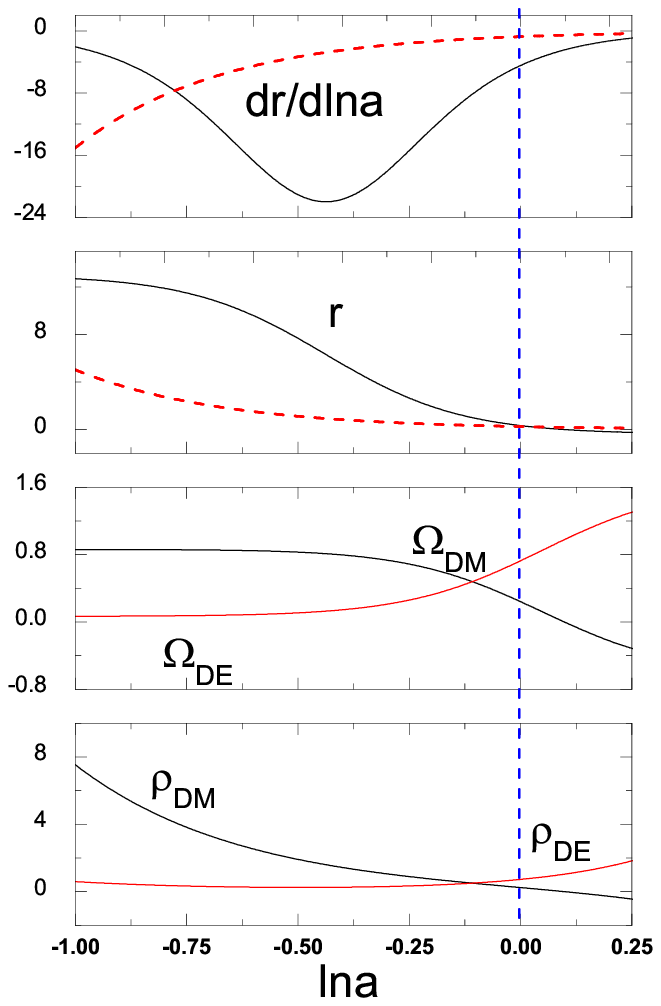}\\
   (a)$\omega_I$&(b)$\omega_{II}$&constant $\omega$
  \end{tabular}
\end{center}
\caption{These figures show clearly the behaviors of different
dark sectors during the late time evolution of universe with
best-fitted data for the coupling $
H(\delta_1\rho_{DE}+\delta_2\rho_{DM})$. The red dotted line
denotes the noninteracting case.The left one corresponds to EoS
$\omega_I$,the middle one is EoS $\omega_{II}$ and the right one
is constant EoS }\label{figGN}
\end{figure}

If the coupling is in linear combination of energy densities of DE
and DM ,we can see easily from Fig~\ref{figGN},the coincidence
problem gets even worse , since $r$ changes even faster than that of
the noninteracting case.Furthermore,we find the negative $\rho_{DM}$
in the future for all the cases,which is unphysical.

\begin{center}
\begin{tabular}{c|c|cccccccc}
 \hline
  \hline
   Coupling & EoS & &$\Omega_b^0$ & $\omega_0$ & $\omega_1$ & $\Omega_{DM}^0$ & $h_0$ & $\delta1$ & $\delta2$ \\
  \hline
  $H(\delta_1\rho_{DE}$ &$\omega_I$ &Mean &$0.043_{- 0.006}^{+ 0.007}$ & $10.97_{- 9.42}^{+ 8.66}$ & $-473.12_{- 611.05}^{+ 220.46}$ & $0.30_{- 0.05}^{+ 0.05}$ & $0.72_{- 0.05}^{+ 0.06}$ & $-13.33 _{- 4.13}^{+ 4.57 }$ & $0.39_{- 0.05}^{+ 0.05}$  \\
  $+\delta_2\rho_{DM})$& & Best-fitted & $0.037 $ & $1.41$ & $ -42.10$ & $0.27$ & $0.77$ & $-4.26 $ & $0.30$  \\
\hline
   &$\omega_{II}$&Mean & $0.048_{- 0.009}^{+ 0.012}$& $1.55_{- 0.34}^{+ 1.64}$& $-188.36_{- 82.24}^{+ 39.10 }$ & $0.23_{- 0.03}^{+ 0.03}$ & $0.69_{- 0.09}^{+ 0.08}$ & $-13.67_{- 5.90}^{+ 10.10}$ & $0.56_{- 0.08}^{+ 0.09}$ \\

&  & Best-fitted &$0.043 $& $0.43 $& $-34.65  $ & $0.24  $ & $0.71$ & $-4.55$ & $0.53$ \\
\hline
  & & &$\Omega_b^0$ & $\omega$  & $\Omega_{DM}^0$ & $h_0$ & $\delta1$ & $\delta2$& \\
  \hline
   &  constant $\omega$ & Mean  & $0.032_{-0.002}^{+0.002}$ & $-2.31_{- 0.61}^{+ 0.39}$ & $0.24_{- 0.02}^{+ 0.02}$ & $0.82_{- 0.03}^{+ 0.03}$ & $-5.56_{-1.87}^{+ 1.33}$& $0.61_{-0.18}^{+0.22}$ & \\
     &              & Best-fitted  & $0.032$ & $-1.40$ & $0.24$ & $0.82$ & $-2.47$ & $0.49$ &  \\
  \hline
  \hline
\end{tabular}
\end{center}
\begin{figure}
\includegraphics[width=2.0in,height=2.0in,angle=-90]{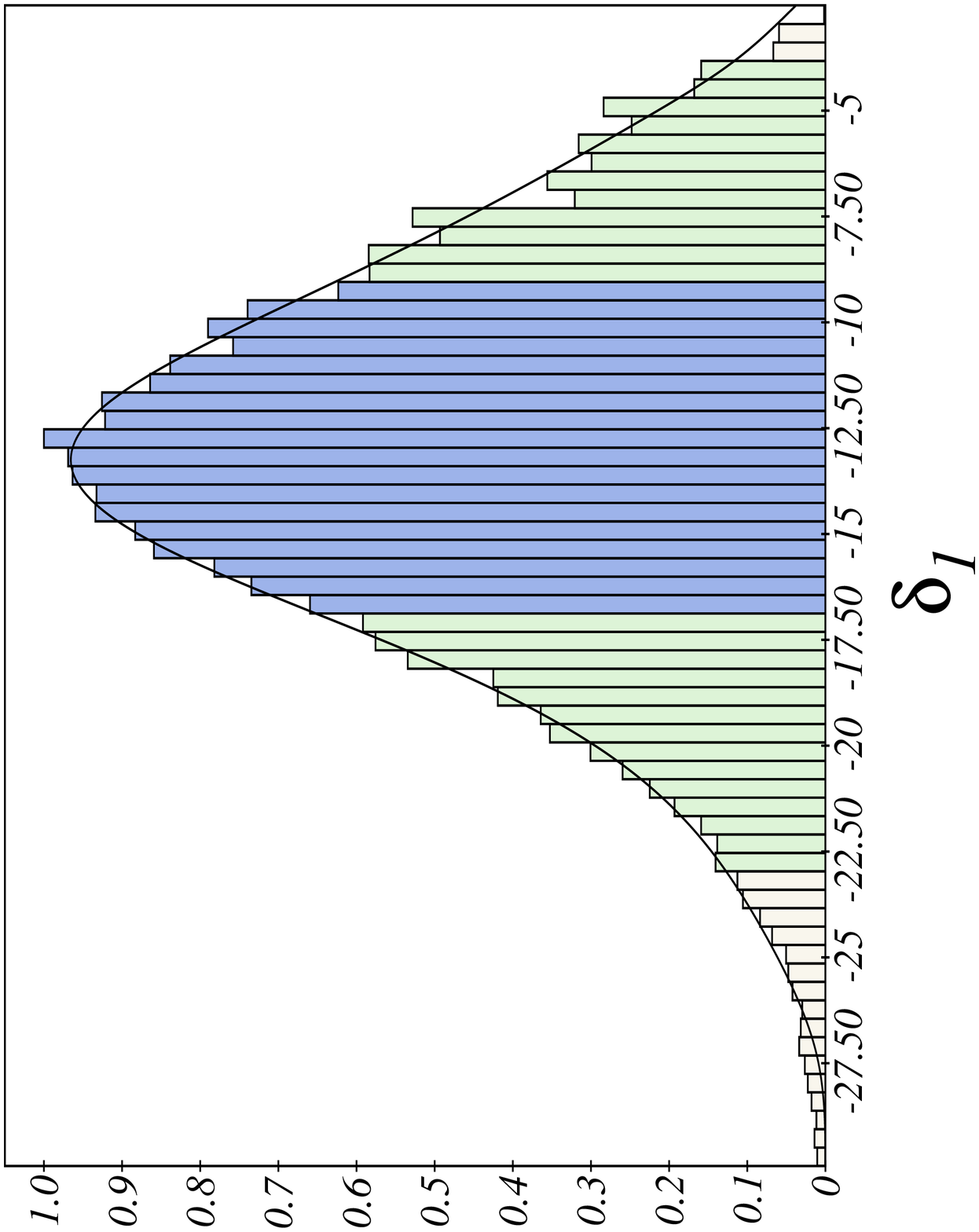}
\includegraphics[width=2.0in,height=2.0in,angle=-90]{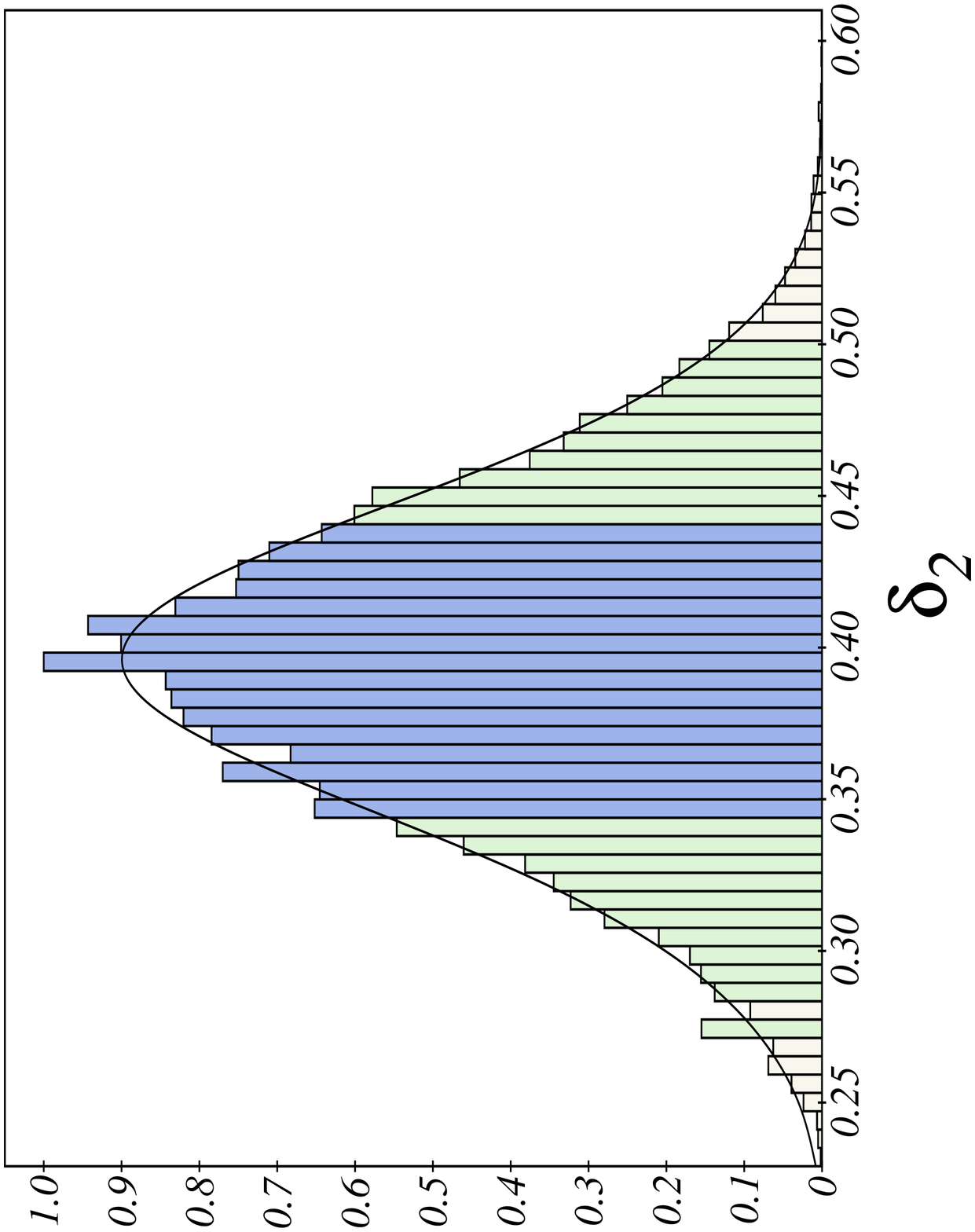}
\caption{The likelihood for coupling parameters when the interaction
is $ H(\delta_1\rho_{DE}+\delta_2\rho_{DM})$}.The left one shows the
likelihood for parameter $\delta_1$ and the right one show the
likelihood for parameter $\delta_2$. The EoS is chosen as $\omega_I$
for these figures. \label{figlikeGN1}
\end{figure}

\begin{figure}
\includegraphics[width=2.0in,height=2.0in,angle=-90]{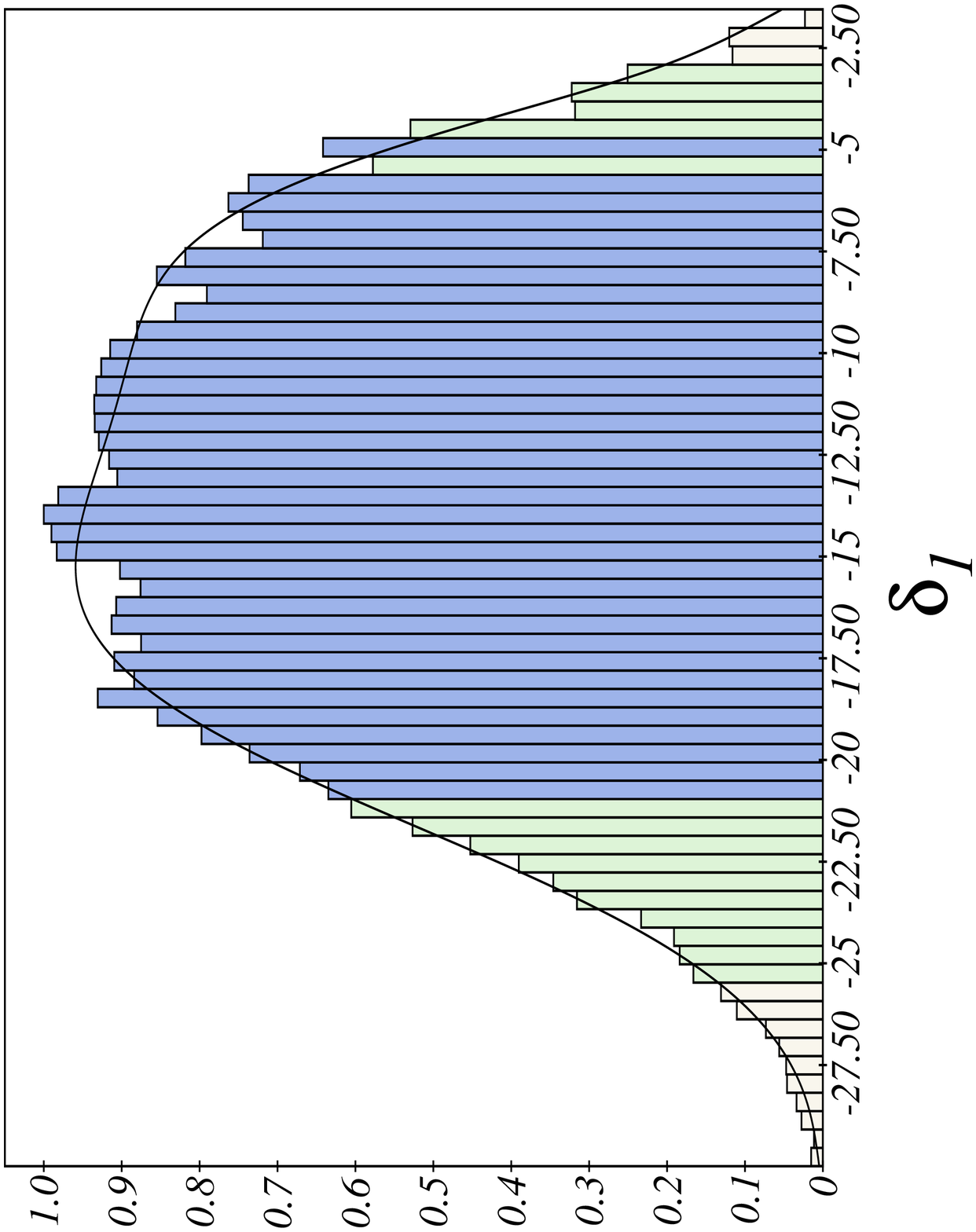}
\includegraphics[width=2.0in,height=2.0in,angle=-90]{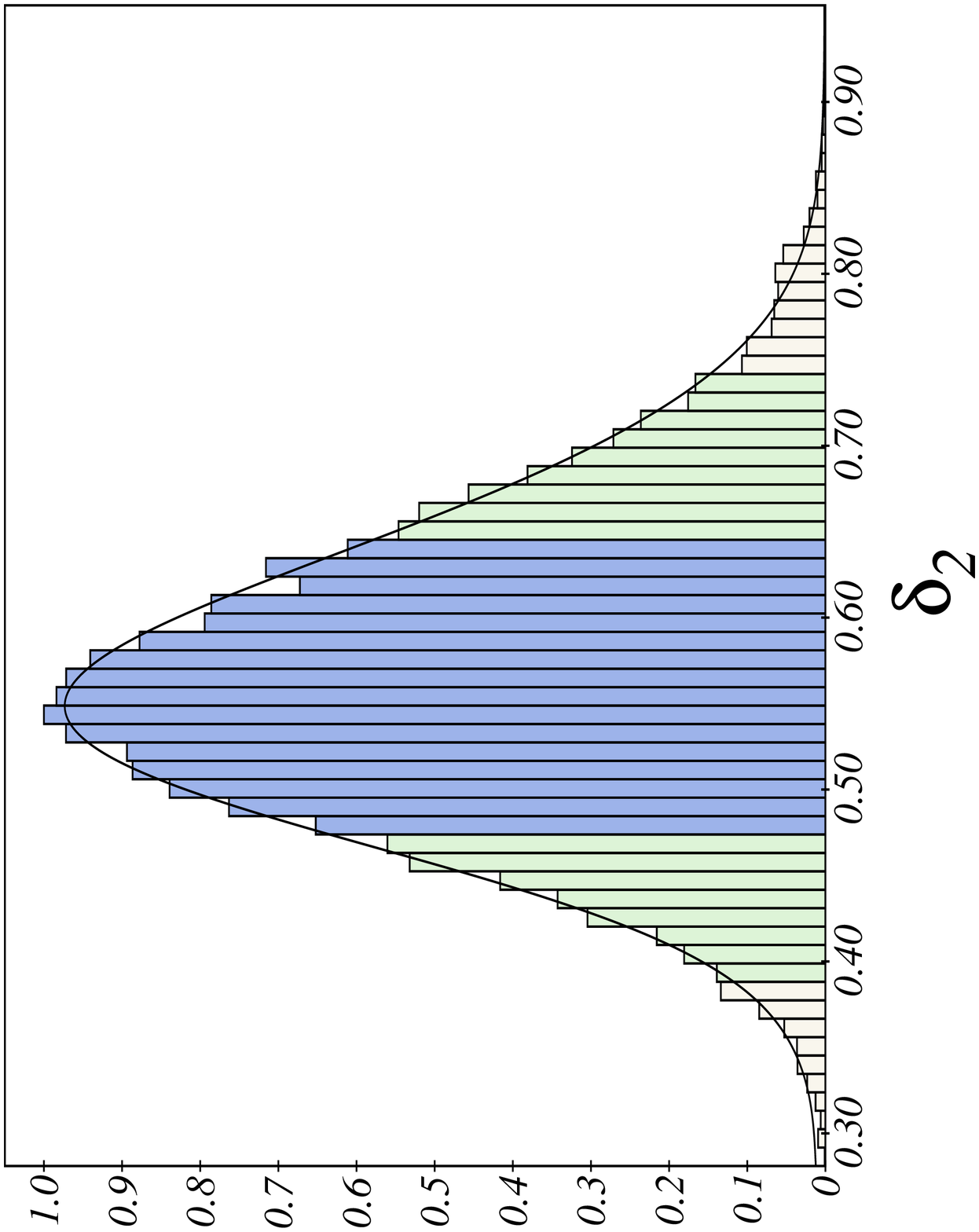}
\caption{The likelihood for coupling parameters when the interaction
is $ H(\delta_1\rho_{DE}+\delta_2\rho_{DM})$}.The left one shows the
likelihood for parameter $\delta_1$ and the right one show the
likelihood for parameter $\delta_2$. The EoS is chosen as
$\omega_{II}$ for these figures. \label{figlikeGN2}
\end{figure}

\subsection{The coupling is proportional to the product of energy densities of DE and DM}
\begin{figure}
\begin{center}
  \begin{tabular}{ccc}
   \includegraphics[width=2in,height=2in]{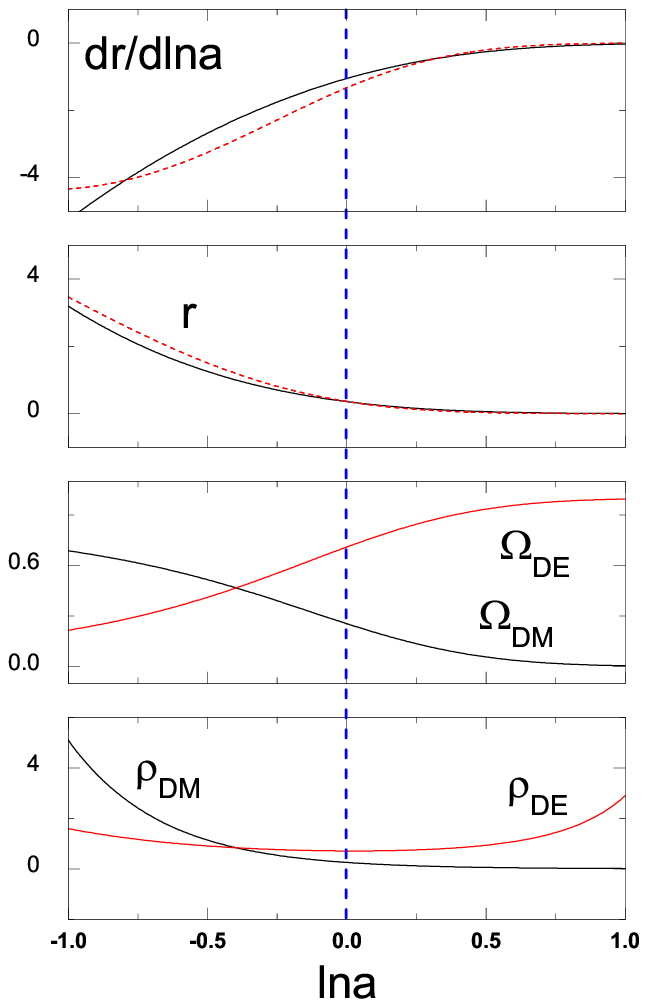}&
   \includegraphics[width=2in,height=2in]{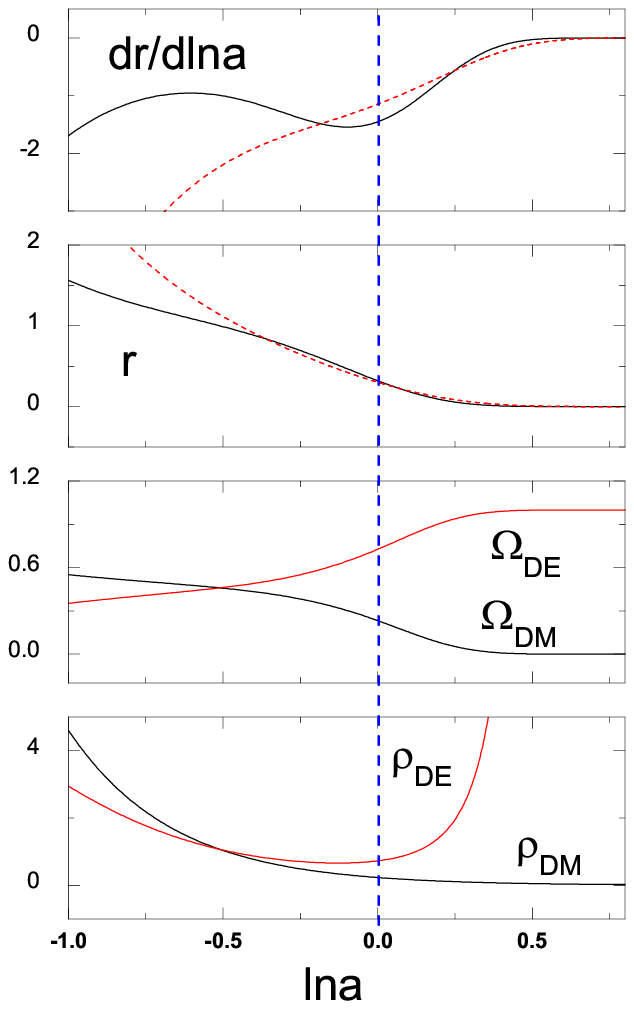}&\includegraphics[width=2in,height=2in]{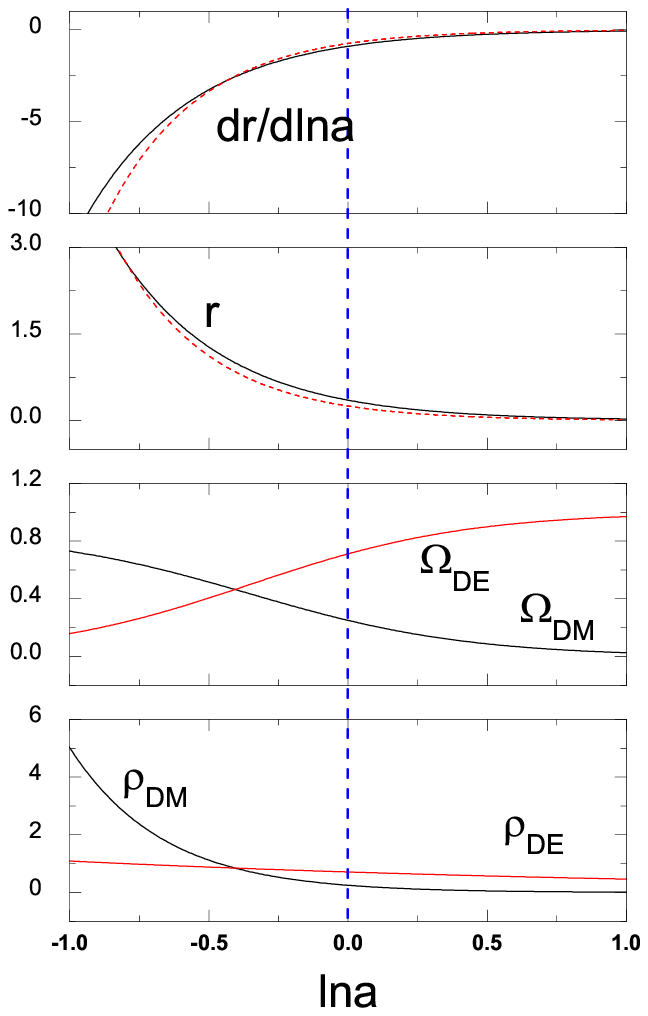}\\
   (a)$\omega_I$&(b)$\omega_{II}$&constant $\omega$
  \end{tabular}
\end{center}
\caption{These figures show clearly the behaviors of different
dark sectors during the late time evolution of universe with
best-fitted data for the interaction $\lambda\rho_{DM}\rho_{DE}$.
The red dotted line denotes the noninteracting case.The left one
corresponds to EoS $\omega_I$,the middle one is EoS $\omega_{II}$
and the right one is constant EoS }\label{figPD}
\end{figure}

In Fig~\ref{figPD},we have shown the results by choosing the
interaction as the product of energy densities of DE and DM
$\omega_I$.Choosing DE EoS $\omega_{I}$,it is clear that with this
kind of interaction between DE and DM, there is a slower change of
$r$ as compared to the noninteracting case. This means that the
period when energy densities of DE and DM are comparable is longer
compared to the noninteracting case. Thus it is not so strange that
we now live in the coincidence state of the universe. In this sense
the coincidence problem is less acute when compared with the case
without interaction. Adopting this kind of interaction form, we have
not met unphysical problems as mentioned above.Choosing DE EoS as
$\omega_{II}$,we see that in most period, $r$ changes slower than
that of the noninteracting case. Only very recently $r$ changes a
bit faster for the interacting case. This will not change the
qualitative behavior of the period when energy densities of DE and
DM are comparable, which is still longer compared to the
noninteracting case. Thus the coincidence problem can become less
serious. Choosing the interaction in the form of the product of
densities of DE and DM, we have positive coupling from the fitting,
which can help to avoid unphysical problems mentioned above.For
constant DE EoS,we see that $r$ changes a bit faster than that of
the noninteracting case near the present epoch, which cannot help to
alleviate the coincidence problem. However the positive coupling
from fitting again avoids unphysical problems for the evolution of
cosmological parameters.
\begin{center}
\begin{tabular}{c|c|ccccccc}
 \hline
 \hline
  Coupling & EoS &  & $\Omega_b^0$ & $\omega_0$ & $\omega_1$ & $\Omega_{DM}^0$ & $h_0$ & $\delta$ \\
   \hline
  $\lambda\rho_{DE}\rho_{DM}$  &$\omega_{I}$  & Mean& $0.036_{-0.004}^{+0.003}$  & $-0.93_{-0.11}^{+0.11}$ &  $0.48_{-0.25}^{+0.16}$ &  $0.25 _{-0.018}^{+0.018}$&  $0.78_{-0.037}^{+0.047}$ &  $0.012_{-0.004}^{+0.004}$ \\

   &  &Best-fitted &$0.036$  & $-0.98$ &  $0.68$ &  $0.25$&  $0.78$ &  $0.0064$\\
  \hline
   &$\omega_{II}$&Mean& $0.040_{-0.003}^{+0.003}$  & $-1.53_{-0.35}^{+0.35}$ &  $4.90_{-2.26}^{+2.13}$ &  $0.24_{-0.018}^{+0.020}$&  $0.74_{-0.027}^{+0.028}$ & $0.018_{-0.007}^{+0.020}$ \\

     & & Best-fitted &$0.041$  & $-1.52$ &  $4.93$ &  $0.23$&  $0.73$ &  $0.016$\\
 \hline
   &  &  & $\Omega_b^0$ & $\omega$  & $\Omega_{DM}^0$ & $h_0$ & $\delta$& \\
   \hline
   &constant &Mean& $0.039_{-0.003}^{+0.002}$  & $-0.84_{-0.09}^{+0.08}$ &  $0.25_{-0.02}^{+0.02}$ &  $0.74_{-0.03}^{+0.03}$&  $0.0063_{-0.0063}^{+0.0074}$ &\\

    &             & Best-fitted &$0.038$  & $-0.85$ &  $0.25$ &  $0.75$&  $0.0025$& \\
  \hline
  \hline
\end{tabular}
\end{center}
\begin{figure}
\includegraphics[width=2.0in,height=2.0in,angle=-90]{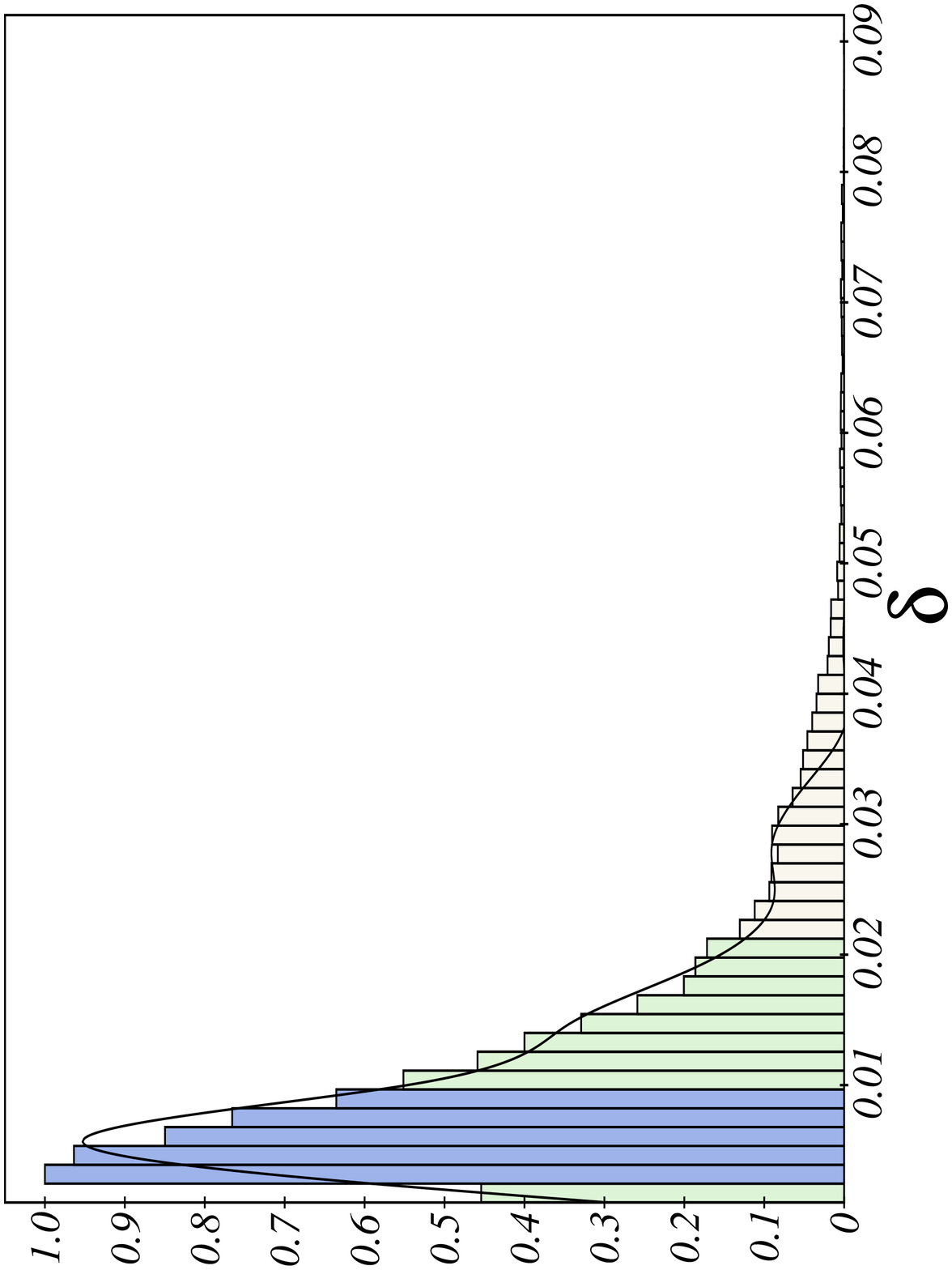}
\includegraphics[width=2.0in,height=2.0in,angle=-90]{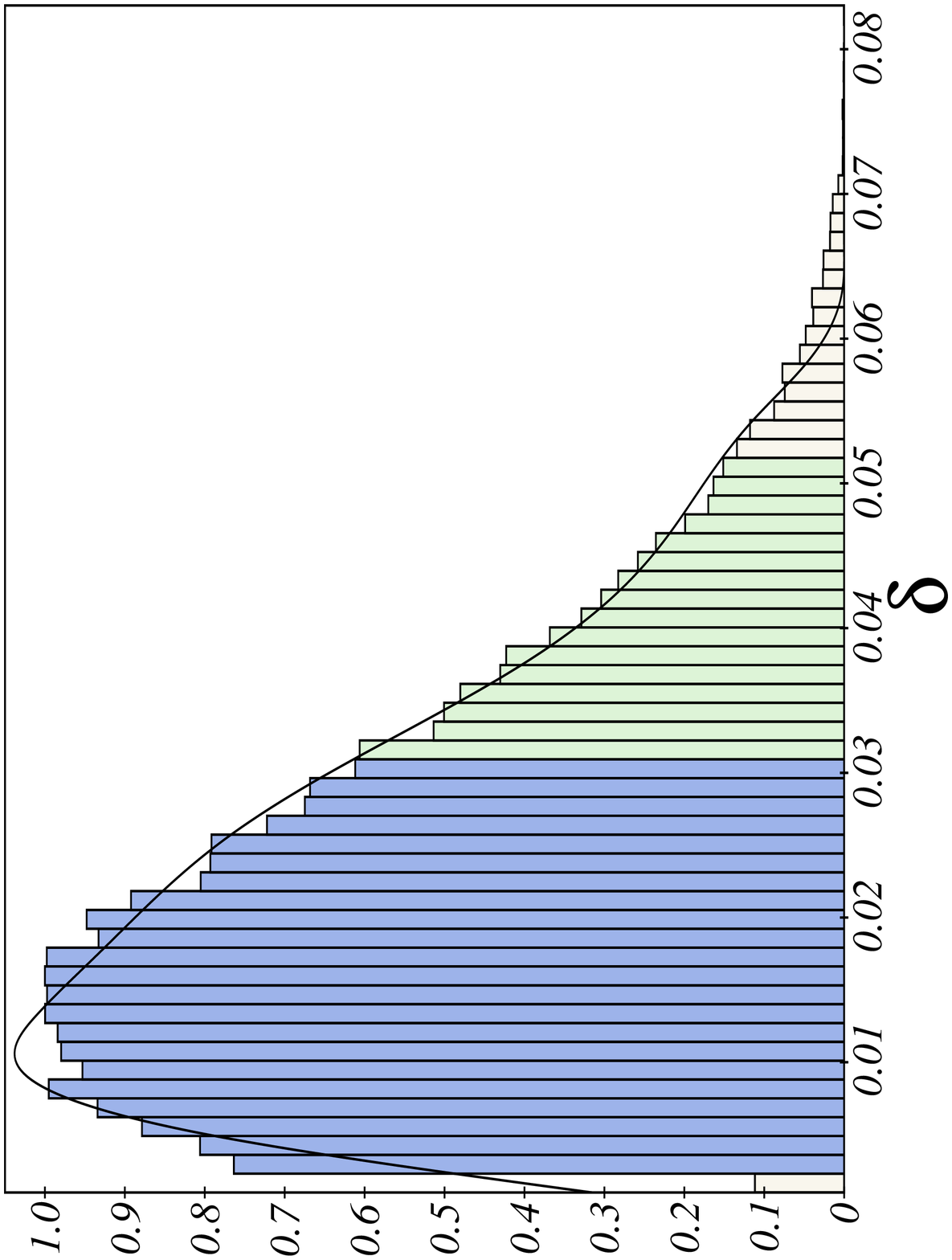}
\caption{The likelihood for coupling parameters when the interaction
is chosen as $\lambda\rho_{DM}\rho_{DE}$}.The left one shows the EoS
$\omega_I$ and the right one is $\omega_{II}$ \label{figlikePD}
\end{figure}

\section{conclusions and discussions}

In this paper we have examined the effects of all possible
phenomenological interactions between DE and DM on cosmological
parameters and their efficiency on solving the coincidence
problem. We have worked with two simple parameterizations of the
DE EoS and the constant DE EoS. Using observational data coming
from the new 182 Gold type Ia supernova samples, the shift
parameter of the Cosmic Microwave Background given by the
three-year Wilkinson Microwave Anisotropy Probe observations, and
the baryon acoustic oscillation measurement from the Sloan Digital
Sky Survey, we have performed a statistical joint analysis of
different forms of phenomenological interaction between DE and DM.
We have found that for the time-dependent DE EoS the model of the
interaction in product of densities of DE and DM is the best model
to alleviate the coincidence problem.  For other interaction
forms, the consequences of DE and DM interaction on cosmological
parameters are very sensitive to the DE EoS. For the DE EoS in the
form of $\omega_I$, choosing the DE and DM interaction in
proportional to the energy density of DE, we found the solution to
solve the coincidence problem. We observed that other forms of the
interaction either bring unphysical problems or cannot help to
alleviate the coincidence problem. For the DE EoS $\omega_{II}$,
we found that except the interaction in terms of the product of
densities of DE and DM, other phenomenological descriptions of
interaction between DE and DM fail to alleviate the coincidence
problem. For the constant DE EoS, none of the phenomenological
models of interaction can help to alleviate the coincidence
problem. It would be interesting to extend our analysis to more
general parameterizations for both the interaction and the DE EoS
available in the literature. Furthermore it is also of interest to
include more observational data to constrain the interaction
between DE and DM and study its consequences on the cosmological
parameters.

\acknowledgments{ This work was partially supported by the NNSF of
China, Shanghai Education Commission, Shanghai Science and
Technology Commission.}


\begin{thebibliography}{0}
\bibitem{1} A. G. Riess {\it et al}, {\it Astron. J.} {\bf 116}, 1009 (1998);
S. Perlmutter {\it et al}, {\it Astrophys. J.} {\bf 517}, 565
(1999); P. de Bernardis {\it et. al.}, {\it Nature} {\bf 404}, 955
(2000).

\bibitem{Perlmutter03}S. Perlmutter {\it et al}, {\it Astrophys. J.} {\bf 598},
102 (2003).

\bibitem{b312} D. J. Eisenstein, {\it et al.}, {\it Astrophys. J.} {\bf 633}, 560 (2005).

\bibitem{WMAP3y} D. N. Spergel {\it et al.},  {\it Astrophys. J. Suppl.} {\bf170},
377 (2007).

\bibitem{b310} A. G. Riess, {\it et al.}, astro-ph/0611572.

\bibitem{2} T. Padmanabhan, {\it Phys. Rept.} {\bf380}, 235 (2003);
P. J. E.Peebles, B. Ratra, {\it Rev. Mod. Phys.}  {\bf75}, 559
(2003); V. Sahni, {\it Lect Notes Phys.} {\bf 653}:141 (2004) and
references therein.

\bibitem{45}L. Amendola, {\it Phys. Rev.} {\bf D62}, 043511 (2000); L. Amendola and C.
Quercellini, {\it Phys. Rev.} {\bf D68}, 023514 (2003); L.
Amendola, S. Tsujikawa and M. Sami, {\it Phys. Lett.} {\bf B632},
155 (2006).

\bibitem{48}D. Pavon, W. Zimdahl, {\it Phys. Lett.} {\bf B628}, 206 (2005).

\bibitem{49} S. Campo, R. Herrera, G. Olivares and D. Pavon, {\it Phys. Rev.} {\bf D74},
023501 (2006); S. Campo, R. Herrera and D. Pavon, {\it Phys. Rev.}
{\bf D71}, 123529 (2005); G. Olivares, F. Atrio-Barandela and D.
Pavon, {\it Phys. Rev.} {\bf D71}, 063523 (2005).

\bibitem{410} G. Olivares, F. Atrio-Barandela and D. Pavon, {\it Phys. Rev.} {\bf D74},
043521 (2006).

\bibitem{411}B. Wang, Y. G. Gong and E. Abdalla, {\it Phys. Lett.} {\bf B624}, 141
(2005).

\bibitem{WLA06}B. Wang, C. -Y. Lin and E. Abdalla, {\it Phys. Lett.} {\bf
B637}, 357 (2006).

\bibitem{412} B. Wang, J. Zang, C.-Y. Lin, E. Abdalla and S. Micheletti, {\it Nucl.
Phys.}  {\bf B778}, 69 (2007).

\bibitem{MPLA} G. Mangano, G. Miele and V. Pettorino, Mod.Phys.Lett.A 18,
831(2003).

\bibitem{413} S. Das, P. S. Corasaniti and J. Khoury, {\it Phys. Rev.} {\bf D73}, 083509
(2006).

\bibitem{414} R. Bean, E. E. Flanagan, M. Trodden, arXiv:0709.1124; R. Bean, E. E. Flanagan, M. Trodden,
arXiv:0709.1128; M. Manera, D. F. Mota arXiv:astro-ph/0504519; N.
J.Nunes, D. F. Mota arXiv:astro-ph/0409481

\bibitem{415} W. Zimdahl, {\it Int. J. Mod. Phys.} {\bf D14}, 2319 (2005).

\bibitem{416} E. Abdalla and B. Wang, {\it Phys. Lett.} {\bf
B651}, 89 (2007).

\bibitem{Feng07}C. Feng, B. Wang, Y. G. Gong, R. K. Su ,
{\it JCAP} {\bf 09}, 005 (2007).


\bibitem{418} Z. K. Guo, N. Ohta and S. Tsujikawa, {\it Phys. Rev.} {\bf D76}, 023508
(2007).

\bibitem{426} O. Bertolami, F. Gil Pedro and M. Le Delliou,  {\it Phys. Lett.} {\bf B654},
165 (2007). O. Bertolami, F. Gil Pedro and M. Le Delliou,
arXiv:0705.3118v1.

\bibitem{427} M. Kesden and M. Kamionkowski, {\it Phys. Rev. Lett.} {\bf 97}, 131303
(2006); M. Kesden and M. Kamionkowski, {\it Phys. Rev.} {\bf D74},
083007 (2006).


\bibitem{Abdalla07}E. Abdalla, L.Raul W. Abramo, L. Sodre Jr.,
B. Wang, arXiv:0710.1198 [astro-ph].

\bibitem{L10} L. Amendola, D. Tocchini-Valentini, Phys. Rev.
D 64 (2001) 043509;  G. W. Anderson, S. M. Carroll,
astro-ph/9711288.

\bibitem{L11} S. Das, P.S. Corasaniti, J. Khoury, Phys. Rev. D 73 (2006) 083509.

\bibitem{L6} W. Zimdahl, D. Pav¡äon, L.P. Chimento, Phys. Lett. B 521 (2001) 133; L.P. Chimento, A.S.
Jakubi, D. Pav¡äon, W. Zimdahl, Phys. Rev. D 67 (2003) 083513; S.
del Campo, R. Herrera, D. Pav¡äon, Phys. Rev. D 70 (2004) 043540;
D. Pav¡äon, W. Zimdahl, Phys. Lett. B 628 (2005) 206.

\bibitem{LL} B. Wang, C.-Y. Lin, D. Pavon, E. Abdalla,
{\it Phys. Lett.} {\bf B662},1 (2008).

\bibitem{226} D. Pavon, B. Wang, arXiv:0712.0565.

\bibitem{Y9} U. Alam, V. Sahni, and A. Starobinsky, JCAP 0406, 008 (2004); Y. G. Gong, Class. Quant.
Grav. 22, 2121 (2005); Y. Wang and M. Tegmark, Phys. Rev. D 71,
103513 (2005); Y. Wang and P. Mukherjee, Astrophys. J. 606, 654
(2004); R. Daly and S. Djorgovski, Astrophys. J. 612, 652 (2004);
U. Alam, V. Sahni, T. Saini, and A. Starobinsky, Mon. Not. Roy.
Astron. Soc. 354, 275 (2004); T. Choudhury and T. Padmanabhan,
Astron. Astrophys. 429, 807 (2005).

\bibitem{GSpergel}  D. N. Spergel, et al., 2007, ApJS, 170, 377.

\bibitem{GRiess}  A. G. Riess, et al., 2006, arXiv:
astro-ph/0611572.

\bibitem{GEisenstein}  D. J. Eisenstein, et al., 2005, ApJ, 633,
560.

\bibitem{GElgar}  O. Elgaroy,   T. Multamaki, 2007,
arXiv:astro-ph/0702343.

\bibitem{GwangY}  Y. Wang,  P. Mukherjee, 2007,
arXiv:astro-ph/0703780.

\bibitem{GFreedman} W. L.  Freedman, et al., 2001, ApJ, 553, 47.

\bibitem{amendola} L. Amendola, G. Campos, R. Rosenfeld,
astro-ph/0610806.

\bibitem{amendola28} L. Amendola, D. Polarski, S. Tsujikawa,
astro-ph/0603703.

\bibitem{cmbeasy} M. Doran, C.M. Muller,
astro-ph/0311311.

\end{thebibliography}
\end{document}